\pdfoutput=1

\documentclass[11pt]{article}

\usepackage[final]{acl}

\usepackage{times}
\usepackage{latexsym}
\usepackage{multirow}
\usepackage{bm}
\usepackage{amsmath}
\usepackage{hyperref} 
\usepackage{microtype}
\usepackage{amsfonts}
\usepackage{graphicx}
\usepackage{subcaption}
\usepackage[most]{tcolorbox}
\tcbuselibrary{skins}
\usepackage{float}

\usepackage[T1]{fontenc}

\usepackage[utf8]{inputenc}

\usepackage{microtype}

\usepackage{inconsolata}

\usepackage{graphicx}

\usepackage{subcaption}
\usepackage{tikz}
\usepackage{booktabs} 

\title{Spark-TTS: An Efficient LLM-Based Text-to-Speech Model with Single-Stream Decoupled Speech Tokens}

\author{
 \textbf{\small Xinsheng Wang\textsuperscript{1,2},}\thanks{Project leader.}
 \textbf{\small Mingqi Jiang\textsuperscript{2,3},}
 \textbf{\small Ziyang Ma\textsuperscript{4,5},}
 \textbf{\small Ziyu Zhang\textsuperscript{6},}
 \textbf{\small Songxiang Liu\textsuperscript{7},}
 \textbf{\small Linqin Li\textsuperscript{3},}
 \textbf{\small Zheng Liang\textsuperscript{4},}  \\
 \textbf{\small Qixi Zheng\textsuperscript{4},}
 \textbf{\small Rui Wang\textsuperscript{3},}
 \textbf{\small Xiaoqin Feng\textsuperscript{3},}
 \textbf{\small Weizhen Bian\textsuperscript{1},}
 \textbf{\small Zhen Ye\textsuperscript{1},}
 \textbf{\small Sitong Cheng\textsuperscript{1},} 
 \textbf{\small Ruibin Yuan\textsuperscript{1},} \\
 \textbf{\small Zhixian Zhao\textsuperscript{6},}
 \textbf{\small Xinfa Zhu\textsuperscript{6},}
 \textbf{\small Jiahao Pan\textsuperscript{1},} 
 \textbf{\small Liumeng Xue\textsuperscript{1,2},} 
 \textbf{\small Pengcheng Zhu\textsuperscript{2,8},} 
 \textbf{\small Yunlin Chen\textsuperscript{3},} \\
 \textbf{\small Zhifei Li\textsuperscript{3},}
 \textbf{\small Xie Chen\textsuperscript{4},}
 \textbf{\small Lei Xie\textsuperscript{6},} 
 \textbf{\small Yike Guo\textsuperscript{1},} 
 \textbf{\small Wei Xue\textsuperscript{1}}\thanks{Corresponding author.}
\\
\small{\textsuperscript{1}}\small{Hong Kong University of Science and Technology}
\small{\textsuperscript{2}}\small{SparkAudio Open Source Community} \\
\small{\textsuperscript{3}}\small{Shanghai Mobvoi Information Technology Co., Ltd}
\small{\textsuperscript{4}}\small{Shanghai Jiao Tong University}
\small{\textsuperscript{5}}\small{Nanyang Technological University} \\
\small{\textsuperscript{6}}\small{ASLP@NPU, Northwestern Polytechnical University}
\small{\textsuperscript{7}}\small{Independent Researcher}
\small{\textsuperscript{8}}\small{Fuxi AI Lab, NetEase, Inc}
\\
 \small{
  \href{w.xinshawn@gmail.com}{w.xinshawn@gmail.com}, \href{ weixue@ust.hk}{ weixue@ust.hk}
 } \\
}

\begin{document}
\maketitle
\begin{abstract}

Recent advancements in large language models (LLMs) have driven significant progress in zero-shot text-to-speech (TTS) synthesis. However, existing foundation models rely on multi-stage processing or complex architectures for predicting multiple codebooks, limiting efficiency and integration flexibility. To overcome these challenges, we introduce Spark-TTS, a novel system powered by BiCodec, a single-stream speech codec that decomposes speech into two complementary token types: low-bitrate semantic tokens for linguistic content and fixed-length global tokens for speaker attributes. This disentangled representation, combined with the Qwen2.5 LLM and a chain-of-thought (CoT) generation approach, enables both coarse-grained control (e.g., gender, speaking style) and fine-grained adjustments (e.g., precise pitch values, speaking rate). To facilitate research in controllable TTS, we introduce VoxBox, a meticulously curated 100,000-hour dataset with comprehensive attribute annotations. Extensive experiments demonstrate that Spark-TTS not only achieves state-of-the-art zero-shot voice cloning but also generates highly customizable voices that surpass the limitations of reference-based synthesis. Source code, pre-trained models, and audio samples are available at~\url{https://github.com/SparkAudio/Spark-TTS}.

\end{abstract}

\section{Introduction}

Recent advances in speech tokenization have revolutionized text-to-speech (TTS) synthesis by bridging the fundamental gap between continuous speech signals and discrete token-based large language models (LLMs)~\cite{anastassiou2024seed, zhu2024unistyle, wang2024streamvoice+}. Through sophisticated quantization techniques, particularly Vector Quantization (VQ)~\cite{van2017neural} and Finite Scalar Quantization (FSQ)~\cite{mentzer2023finite}, codec-based LLMs have emerged as the predominant paradigm for zero-shot TTS. The integration of extensive training data with large-scale model architectures has enabled these systems to achieve unprecedented levels of naturalness, often rendering synthetic speech indistinguishable from human speech~\cite{anastassiou2024seed, du2024cosyvoice2, chen2024f5, ye2024flashspeech}.

\begin{figure}
    \centering
    \includegraphics[width=\linewidth]{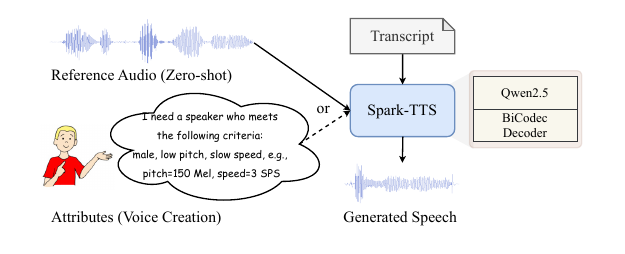}
    \caption{Spark-TTS enables zero-shot voice cloning from reference audio while also generating new speakers through coarse- or fine-grained attribute control. The final waveform is directly reconstructed from the predicted speech tokens using BiCodec’s decoder.}
    \label{fig:illustration}
\end{figure}

Despite the remarkable progress in LLM-based zero-shot TTS, several fundamental challenges persist. Current codec-based TTS architectures exhibit significant complexity, requiring either dual generative models~\cite{wang2023neural,anastassiou2024seed} or intricate parallel multi-stream code prediction mechanisms~\cite{kreukaudiogen,le2024stack} that deviate substantially from conventional text LLM frameworks. This divergence stems from inherent limitations in existing audio codecs - while semantic tokens provide compactness, they necessitate additional models for acoustic feature prediction~\cite{du2024cosyvoice,huang2023repcodec} and lack integrated timbre control capabilities. Acoustic tokens, meanwhile, rely on complex codebook architectures like group-VQ~\cite{defossez2022high,van2017neural}. The field also struggles with the creation of novel voices, as current systems are predominantly limited to reference-based generation~\cite{zhang2023speak,chen2024vall}, lacking the capability to synthesize voices with precisely specified characteristics. This limitation is further compounded by insufficient granularity in attribute control, especially for fine-grained characteristics such as pitch modulation, despite recent advances in instruction-based generation~\cite{du2024cosyvoice2}. Furthermore, the prevalent use of proprietary datasets in current research creates significant challenges for standardized evaluation and meaningful comparison of methods~\cite{anastassiou2024seed,ye2024flashspeech}. These limitations collectively underscore the need for a unified approach that can simplify architecture, enable flexible voice creation with comprehensive attribute control, and establish reproducible benchmarks through open data resources.

To address these fundamental limitations, we introduce Spark-TTS, a unified system that achieves zero-shot TTS with comprehensive attribute control through a single codec LLM, maintaining architectural alignment with conventional text LLMs. In addition, we present VoxBox, a meticulously curated and annotated open-source speech dataset that establishes a foundation for reproducible research in speech synthesis. Specifically, we introduce BiCodec, a novel tokenization framework that preserves the efficiency of semantic tokens while enabling fine-grained control over timbre-related attributes. BiCodec achieves this through combining low-bitrate semantic tokens with fixed-length global tokens, effectively capturing both linguistic content and time-invariant acoustic characteristics. Building upon BiCodec, we leverage Qwen2.5~\cite{yang2024qwen2} through targeted fine-tuning, seamlessly integrating TTS capabilities within the text LLM paradigm. To enable comprehensive voice control, we implement a hierarchical attribute system combining coarse-grained labels (gender, pitch, speaking speed) with fine-grained numerical values, orchestrated through a chain-of-thought (CoT) prediction framework. 

Our primary contributions encompass:

\begin{itemize}
    \item \textbf{New Tokenization}: We present BiCodec, a unified speech tokenization that generates a hybrid token stream combining semantic and global tokens. This approach maintains linguistic fidelity while enabling sophisticated attribute control through LM-based mechanisms.
    \item \textbf{Coarse- and Fine-Grained Voice Control}: Spark-TTS implements a comprehensive attribute control system that seamlessly integrates both categorical and continuous parameters within a text LLM-compatible architecture. As demonstrated in Fig.~\ref{fig:illustration}, this innovation transcends traditional reference-based approaches to zero-shot TTS.
    \item \textbf{Benchmark Dataset}: We introduce VoxBox, a rigorously curated 100,000-hour speech corpus, developed through systematic data collection, cleaning, and attribute annotation. This resource establishes a standardized benchmark for TTS research and evaluation.

\end{itemize}


\section{Related Work}

\subsection{Single-Stream Speech Tokenizer}

Early single-stream speech tokenizers primarily focused on extracting semantic tokens~\cite{huang2023repcodec, du2024cosyvoice, tao2024toneunit}. While pure semantic tokens enable low-bitrate encoding, they necessitate an additional acoustic feature prediction module in semantic token-based speech synthesis~\cite{du2024cosyvoice, du2024cosyvoice2}. 

Recently, single-stream-based acoustic tokenization has gained considerable attention~\cite{xin2024bigcodec, wu2024ts3}. WavTokenizer~\cite{ji2024wavtokenizer} employs a convolution-based decoder to improve reconstruction quality, while X-codec2~\cite{ye2025llasa} enlarges the code space with FSQ. 
Instead of following a pure encoder-VQ-decoder paradigm, decoupling speech content has proven effective in reducing bitrate using a single codebook~\cite{li2024single, zheng2024freecodec}. 

Among these methods, TiCodec~\cite{ren2024fewer} is the most similar to our approach in handling global information. However, unlike TiCodec, the proposed BiCodec employs semantic tokens as its time-variant tokens. Instead of using group GVQ~\cite{ren2024fewer}, we propose a novel global embedding quantization method based on FSQ with learnable queries and a cross-attention mechanism. This approach enables the generation of a relatively longer token sequence, offering a more expressive and flexible representation.

\subsection{LLM-based Zero-Shot TTS}

Prevalent codec LLMs zero-shot TTS predominantly fall into two categories. The first type involves predicting single-stream codes using LLMs, followed by the generation of codes enriched with detailed acoustic or continuous semantic features through another LLM~\cite{ zhang2023speak, chen2024vall, wang2024speechx} or generative diffusion models~\cite{anastassiou2024seed, casanova2024xtts}. The second type involves predicting multi-stream codes using carefully designed parallel strategies~\cite{le2024stack, copet2024simple} or masked generative patterns~\cite{garcia2023vampnet, ziv2024masked, li2024masksr}.

By leveraging the single-stream tokens produced by the proposed BiCodec, Spark-TTS simplifies the modeling of speech tokens within an LLM framework that is fully unified with text LLMs. The most comparable work is the concurrent TTS model Llasa~\cite{ye2025llasa}, which employs an FSQ-based tokenizer to encode speech into single-stream codes with a codebook size of 65,536, followed by LLaMA~\cite{touvron2023llama} for speech token prediction. In contrast, Spark-TTS extends beyond zero-shot TTS by integrating speaker attribute labels, enabling controllable voice creation. Additionally, Spark-TTS achieves higher zero-shot TTS performance while using fewer model parameters, enhancing both efficiency and flexibility.

\section{BiCodec}

\begin{figure}[tp]
    \centering
    \includegraphics[width=0.95\linewidth]{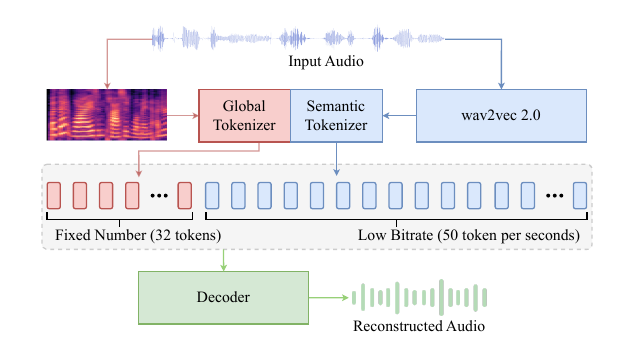}
    \caption{Illustration of the BiCodec. The Global Tokenizer processes the Mel spectrogram to produce global tokens with fixed length, while the Semantic Tokenizer adopts features from wav2vec~2.0 to produce 50~TPS semantic tokens. The decoder reconstructs the waveform from the generated tokens. The detailed structure of BiCodec is provided in Appendix~\ref{append:bicodec}.}
    \label{fig:bicodec}
    \vspace{-15pt}
\end{figure}

To achieve both the compact nature and semantic relevance of semantic tokens, while also enabling acoustic attribute control within an LM, we propose BiCodec, which discretizes input audio into: (i) Semantic tokens at 50 tokens per second (TPS), capturing linguistic content, and (ii) Fixed-length global tokens, encoding speaker attributes and other global speech characteristics.

\subsection{Overview}

As shown in Fig.~\ref{fig:bicodec}, BiCodec includes a Global Tokenizer and a Semantic Tokenizer. The former extracts global tokens from the Mel spectrogram of input audio. The latter uses features from wav2vec~2.0~\cite{baevski2020wav2vec} as input to extract semantic tokens.

The BiCodec architecture follows a standard VQ-VAE encoder-decoder framework, augmented with a global tokenizer. The decoder reconstructs discrete tokens back into audio. For an input audio signal $\bm{x} \in {[-1, 1]}^T$, with sample number of $T$, BiCodec functions as follows:
\begin{equation}
\begin{aligned}
    & \bm{z}    = E_s(F(\bm{x})),  \bm{g} = E_g(\text{Mel}(\bm{x})), \\
    & \bm{g}_f  = \text{CrossAttention}(\bm{g}, \bm{h}), \\
    & \bm{z}_q  = Q_s(\bm{z}),  \bm{g}_q = Q_g(\bm{g}_f), \\
    & \bm{\hat{x}}  = G(\bm{z}_q, A_g(\bm{g}_q)),
\end{aligned}
\end{equation}
where $E_s(\cdot)$ is the encoder of the semantic tokenizer, $F(\cdot)$ is the pre-trained wav2vec~2.0\footnote{\scriptsize{\url{https://huggingface.co/facebook/wav2vec2-large-xlsr-53}}}, $E_g(\cdot)$ is the encoder of the global tokenizer, $\text{Mel}(\cdot)$ is to extract Mel spectrogram from $\bm{x}$, $\bm{h}$ is a sequence of learnable queries matching the length of the final global token sequence, $Q_s(\cdot)$ is a quantization layer with VQ, $Q_g(\cdot)$ is a quantization layer with FSQ, $A_g(\cdot)$ is an aggregation module with a pooling layer, and $G(\cdot)$ is the decoder that reconstructs the time-domain signal $\bm{\hat{x}}$. 

\subsection{Model Structure}

\begin{figure*}[htp]
    \centering
    \includegraphics[width=0.9\linewidth]{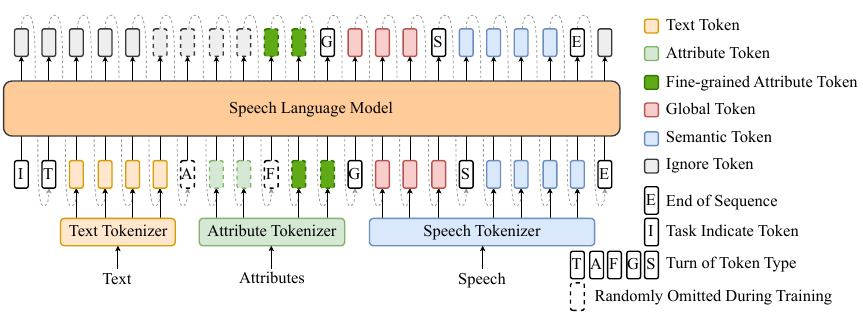}
    \caption{Speech language model of Spark-TTS. During inference, if the input contains attribute tokens representing gender, pitch level, and speed level, the model can predict the corresponding fine-grained attribute tokens, global tokens, and semantic tokens without requiring reference audio in a CoT manner. Otherwise, global tokens can be derived from the reference audio for zero-shot TTS.}
    \label{fig:llm}
    \vspace{-15pt}
\end{figure*}

\textbf{Encoder and Decoder}
The encoder of the semantic tokenizer $E_s$ and the decoder $G$ are fully convolutional neural networks built with ConvNeXt~\cite{liu2022convnet} blocks. To effectively capture semantic information, based on the relationship between different layer features of wav2vec 2.0 (XLSR-53) and semantics~\cite{pasad2023comparative}, we select features from the 11th, 14th, and 16th layers, averaging them to obtain the semantic feature, which serves as the input for the semantic tokenizer. The features from the first two layers show a strong correlation with words, while the features from the 16th layer exhibit the strongest correlation with phonemes.

The global tokenizer's encoder, $E_g$, uses the ECAPA-TDNN architecture~\cite{desplanques2020ecapa} following the implementation by Wespeaker~\cite{wang2023wespeaker} up to the final pooling layer. After encoding, the global tokenizer extracts a fixed-length sequence representation $\bm{g}_f$ using a cross-attention mechanism with a set of learnable queries.

\textbf{Quantization} The semantic tokenizer employs single-codebook vector quantization for quantization. Inspired by DAC~\cite{kumar2024high}, we use factorized codes to project the encoder's output into a low-dimensional latent variable space prior to quantization.

Considering that the global tokenizer requires a set of discrete tokens to represent time-independent global information, FSQ is employed rather than VQ to mitigate the potential risk of training collapse associated with VQ. Details about the model structure can be seen in Appendix~\ref{append:bicodec}. 

\subsection{Training objective}
\textbf{Loss Functions}
BiCodec is trained end-to-end employing a Generative Adversarial Network (GAN) methodology~\cite{goodfellow2020generative} to minimize reconstruction loss, together with L1 feature matching loss (via
discriminators)~~\cite{kumar2019melgan, kumar2024high} while simultaneously optimizing the VQ codebook.

Following~\cite{kumar2024high}, we compute the frequency domain reconstruction loss using L1 loss on multi-scale mel-spectrograms.
Multi-period discriminator~\cite{kong2020hifi, engel2020ddsp, gritsenko2020spectral} and multi-band multi-scale STFT discriminator~\cite{kumar2024high} are used for waveform discrimination and frequency domain discrimination, respectively.

VQ codebook learning incorporates both a codebook loss and a commitment loss. Following the approach in~\cite{xin2024bigcodec}, the codebook loss is calculated as the L1 loss between the encoder output and the quantized results, employing stop-gradients. Additionally, the straight-through estimator~\cite{bengio2013estimating} is used to enable the backpropagation of gradients.

To ensure training stability, in the initial stages, the global embedding derived from the averaged $\bm{g}_q$ is not integrated into the decoder. Instead, this embedding is obtained directly from the pooling of $\bm{g}_f$. Meanwhile, the FSQ codebook is updated using an L1 loss between embedding obtained from $\bm{g}_f$ and that from $\text{pool}(\bm{g}_q)$. As training progresses and stabilizes, this teacher-student form will be omitted after a specific training step.

To further ensure semantic relevance, following X-Codec~\cite{ye2024codec}, a wav2vec 2.0 reconstruction loss is applied after quantization, with ConvNeXt-based blocks serving as the predictor.

\section{Language Modeling of Spark-TTS}
\subsection{Overview}

As illustrated in Fig.~\ref{fig:llm}, the Spark-TTS speech language model adopts a decoder-only transformer architecture, unified with a typical textual language model. We employ the pre-trained textual LLM Qwen2.5-0.5B\footnote{\scriptsize{\url{https://huggingface.co/Qwen/Qwen2.5-0.5B-Instruct}}}~\cite{yang2024qwen2} as the backbone of the speech language model. Unlike CosyVoice2~\cite{du2024cosyvoice}, Spark-TTS does not require flow matching to generate acoustic features. Instead, BiCodec’s decoder directly processes the LM’s output to produce the final audio, significantly simplifying the textual LLM-based speech generation pipeline.

In addition to zero-shot TTS, Spark-TTS supports voice creation using various attribute labels. During inference, if attribute labels for gender, pitch level, and speed level are provided, the language model can predict fine-grained pitch values, speed values, global tokens, and semantic tokens through a chain-of-thought (CoT) manner. If no attribute labels are provided, global tokens are extracted from the reference audio, enabling zero-shot TTS.

\subsection{Tokenizer}
\textbf{Text Tokenizer} Similar to textual LLMs, Spark-TTS employs a byte pair encoding (BPE)-based tokenizer to process raw text. Here, we adopt the Qwen2.5 tokenizer~\cite{yang2024qwen2}, which supports multiple languages. 

\textbf{Attribute Tokenizer} To enable voice creation based on speech attributes, Spark-TTS encodes attribute information at two levels: (i) \emph{Coarse-Grained}: Attribute labels representing high-level speech characteristics, including gender, pitch (categorized into five discrete levels), and speed (categorized into five discrete levels); (ii) \emph{Fine-Grained}: Attribute values enabling precise control over pitch and speed, which are quantized by rounding to the nearest integer during tokenization.

\textbf{Speech Tokenizer} The speech tokenizer consists of a global tokenizer and a semantic tokenizer. Using both global and semantic tokens, the BiCodec decoder reconstructs the waveform signal.

\subsection{Training Objective}
The decoder-only language model is trained by minimizing the negative log-likelihood of token predictions. Let $\mathcal{T}$ represent the tokenized textual prompt and $\mathcal{G}$ denote the global speech token prompt; the optimization for zero-shot TTS is defined as follows:
\vspace{-10pt}
\begin{equation}
    \mathcal{L}_{zst} = -\sum_{t=1}^{T_o} \log P(o_t | \mathcal{T}, \mathcal{G}, \bm{o}_{<t}; \theta_{LM}),
\end{equation}
where $\bm{o} \in  \mathbb{N}^T_o$ represents the semantic tokens to be predicted in the zero-shot TTS scenario, and $\theta_{LM}$ denotes the parameters of the language model. 

For the case of voice creation, the optimization is defined as follows:
\begin{equation}
    \mathcal{L}_{control} = -\sum_{t=1}^{T_c} \log P(c_t | \mathcal{T}, \mathcal{A}, \bm{c}_{<t}; \theta_{LM}),
\end{equation}
where $\mathcal{A}$ represents the attribute label prompt, and the output $\bm{c}$ encompasses $\mathcal{F}$, $\mathcal{G}$, and $\mathcal{S}$. Here, $\mathcal{F}$ denotes the fine-grained attribute value prompt, and $\mathcal{S}$ is speech semantic tokens.
 
In practice, $\mathcal{L}_{zst}$ and $\mathcal{L}_{control}$ are mixed during training. Specifically, each audio example is structured into two training samples according to $\mathcal{L}_{zst}$ and $\mathcal{L}_{control}$ respectively.

\section{VoxBox}

\subsection{Overview}
To facilitate voice creation and establish a fair comparison benchmark for future research, we introduce VoxBox, a well-annotated dataset for both English and Chinese. All data sources in VoxBox originate from open-source datasets, ensuring broad accessibility. To enhance data diversity, we collect not only common TTS datasets, but also datasets used for speech emotion recognition. Each audio file in VoxBox is annotated with gender, pitch, and speed. 
Additionally, we also perform data cleaning on datasets with lower text quality. After data cleaning, VoxBox comprises 4.7 million audio files, sourced from 29 open datasets, totaling 102.5k hours of speech data. Details about VoxBox and the source datasets can be found in Appendix~\ref{ap.sec:voxbox}.

\subsection{Clean and Annotation}
\textbf{Gender Annotation} Given the strong performance of pre-trained WavLM in speaker-related tasks~\cite{li2024scdnet}, we fine-tune the WavLM-large model for gender classification using datasets that contain explicit gender labels (detailed in Appendix~\ref{sc:gender_data_ap}). Our fine-tuned model achieves 99.4\% accuracy on the AISHELL-3 test set. We then use this gender classification model to annotate datasets previously lacking gender labels.

\textbf{Pitch Annotation} 
We extract the average pitch value from each audio clip using PyWorld\footnote{\scriptsize{\url{https://pypi.org/project/pyworld/}}}, rounding it to the nearest integer to obtain fine-grained pitch value tokens.
For the definition of pitch levels, we first convert the average pitch of each audio clip to the Mel scale. We then conduct a statistical analysis of all Mel scale pitch for all males and females separately. Based on the 5th, 20th, 70th, and 90th percentiles, we establish boundaries for five pitch levels: very low, low, moderate, high, and very high (detailed in Appendix~\ref{ap.sec.pitch_class}). 

\textbf{Speed Annotation} Compared to character-based~\cite{vyas2023audiobox}, word-based~\cite{ji2024textrolspeech}, or phoneme-based~\cite{lyth2024natural} speaking rate calculations, syllable-based measurements provide a more direct correlation with speaking rate.
Here, we initially apply Voice Activity Detection (VAD) to eliminate silent segments at both ends. Subsequently, we calculate the syllables per second (SPS), which is then rounded to the nearest integer to serve as the fine-grained speed value token. Using the 5th, 20th, 80th, and 95th percentiles, we establish boundaries for five distinct speed levels: very slow, slow, moderate, fast, and very fast (detailed in Appendix~\ref{ap.sec.pitch_class}). 

\textbf{Data Cleaning} For datasets exhibiting lower text quality, we conduct an additional cleaning process. Specifically, for Emilia~\cite{he2024emilia}, the original transcripts were obtained using the Whisper-based (ASR) system~\cite{radford2023robust}, employing the whisper-medium model, which occasionally resulted in inaccuracies. To address this, we employ another ASR model, FunASR~\cite{gao2023funasr}\footnote{\scriptsize {ZH: \url{https://huggingface.co/funasr/paraformer-zh}} \\
\scriptsize{EN: \url{https://huggingface.co/FunAudioLLM/SenseVoiceSmall}}}, to re-recognize the audio. We then use the original scripts as ground truth to calculate the Word Error Rate (WER) and excluded samples with a WER exceeding 0.05. 
For the MLS-English, LibriSpeech, LibriTTS-R, and datasets originally designed for emotion recognition, we employ the whisper-large-v3\footnote{\scriptsize{\url{https://huggingface.co/openai/whisper-large-v3}}} model for speech recognition, comparing the recognition results with the original scripts. Samples exhibiting insertions or deletions are excluded from the dataset.

\section{Experiments}

\subsection{Implementation Details}
BiCodec is trained on the full training set of the LibriSpeech dataset, comprising 960 hours of English speech data. Additionally, we include 1,000 hours of speech data from both Emilia-CN and Emilia-EN, bringing the total training data to approximately 3,000 hours. All audio samples are resampled to 16 kHz. The global token length is set as 32.
For optimization, we use the AdamW optimizer with moving average coefficients coefficients $\beta_1 = 0.8$ and $\beta_2 = 0.9$. The model converges within approximately 800k training steps using a batch size with 614.4 seconds of speech.
 
The Spark-TTS language model is trained using the entire VoxBox training set. If a dataset lacks predefined train/test splits, we use the entire processed dataset for training. The training employs the AdamW optimizer with $\beta_1 = 0.9$ and $\beta_2 = 0.96$. The model undergoes training over 3 epochs, using a batch size of 768 samples.

\subsection{Reconstruction Performance of BiCodec}

\begin{table*}[htb]
\centering
\small
\caption{Comparisons of various codec models for speech reconstruction on the LibriSpeech test-clean dataset. Detailed information about these models can be found in Appendix~\ref{sc:codec_compare_append}.}
\setlength{\tabcolsep}{2mm}
\begin{tabular}{l|cccc|ccccc}
\toprule
Model           & \begin{tabular}[c]{@{}c@{}}Codebook \\ Size\end{tabular} & Nq & \begin{tabular}[c]{@{}c@{}}Token Rate\\ (TPS)\end{tabular} & \begin{tabular}[c]{@{}c@{}}Bandwidth\\ (bps)\end{tabular} & STOI$\uparrow$ & \begin{tabular}[c]{@{}c@{}}PESQ\\ NB$\uparrow$\end{tabular} & \begin{tabular}[c]{@{}c@{}}PESQ\\ WB$\uparrow$\end{tabular} & UTMOS$\uparrow$ & SIM$\uparrow$  \\ \midrule
Encodec         & 1024       & 8  & 600     & 6000      & 0.94 & 3.17       & 2.75       & 3.07  & 0.89 \\
DAC             & 1024       & 12 & 600     & 6000      & \textbf{0.95} & \textbf{4.15}       & \textbf{4.01}       & 4.00     & \textbf{0.98} \\
Encodec         & 1024       & 2  & 150     & 1500      & 0.84 & 1.94       & 1.56       & 1.58  & 0.6  \\
Mimi            & 2048       & 8  & 100     & 1100      & 0.91 & 2.8        & 2.25       & 3.56  & 0.73 \\
BigCodec        & 8192       & 1  & 80      & 1040      & 0.94 & 3.27       & 2.68       & 4.11  & 0.84 \\
DAC             & 1024       & 2  & 100     & 1000      & 0.73 & 1.4        & 1.14       & 1.29  & 0.32 \\
SpeechTokenizer & 1024       & 2  & 100     & 1000      & 0.77 & 1.59       & 1.25       & 2.28  & 0.36 \\
X-codec         & 1024       & 2  & 100     & 1000      & 0.86 & 2.88       & 2.33       & 4.21  & 0.72 \\ \midrule
WavTokenizer    & 4096       & 1  & 75      & 900       & 0.89 & 2.64       & 2.14       & 3.94  & 0.67 \\
X-codec2        & 65536      & 1  & 50      & 800       & 0.92 & 3.04       & 2.43       & 4.13  & \textbf{0.82} \\
StableCodec     & 15625      & 2  & 50      & 697       & 0.91 & 2.91       & 2.24       & \textbf{4.23}  & 0.62 \\
Single-Codec     & 8192       & 1  & 23.4    & 304       & 0.86 & 2.42       & 1.88       & 3.72  & 0.60  \\ \midrule
BiCodec         & 8192       & 1  & 50      & 650       & \textbf{0.92} & \textbf{3.13}       & \textbf{2.51}       & 4.18  &  0.80    \\ \bottomrule
\end{tabular}
\label{tab:codec_compare}
\vspace{-5pt}
\end{table*}

\textbf{Comparsion with Other Methods} The reconstruction performance of BiCodec compared to other methods is presented in Table~\ref{tab:codec_compare}. As can be seen, within the low-bitrate range (<1 kbps), BiCodec surpasses all methods on most metrics, except for UTMOS, where it ranks second to StableCodec, and SIM, where it ranks second to X-Codec2,  thereby achieving a new state-of-the-art (SOTA) performance.

Notably, BiCodec’s semantic tokens are extracted from wav2vec 2.0 rather than raw audio, resulting in stronger semantic alignment compared to codecs that directly process waveform-based representations. Further experimental results and analyses are provided in Appendix~\ref{sc:codec_exp_append}.

\textbf{Effectiveness of Global Tokenizer}
We first evaluate the optimal length for the global token sequence. As shown in Table~\ref{tab:global_length}, we compare the impact of different sequence lengths on reconstruction quality. The results without FSQ quantization serve as a benchmark reference. Notably, increasing the global token sequence length consistently improves reconstruction quality, with performance approaching the benchmark at the length of 32.

Furthermore, Table~\ref{tab:global_length} compares our proposed quantization method—which incorporates learnable queries and FSQ—against the GVQ-based method introduced by Ren et al.~\cite{ren2024fewer} for time-invariant codes. Our approach demonstrates a substantial performance improvement over the GVQ-based method, highlighting the effectiveness of FSQ with learnable queries in enhancing global token representation.

\begin{table}[htp]
\centering
\small
\caption{Performance of BiCodec with varying global token lengths for reconstruction on the LibriSpeech test-clean dataset, where "w/o" indicates the omission of FSQ-based quantization, and gvq-32 means the global tokenizer is implemented with group VQ. For performance results on the LibriTTS test-clean dataset, refer to Appendix~\ref{sc:codec_exp_append}.}
\setlength{\tabcolsep}{1.8mm}
\begin{tabular}{lccccc}
\toprule
\multicolumn{1}{c}{\begin{tabular}[c]{@{}c@{}}Global\\ Token\end{tabular}} & \multicolumn{1}{c}{STOI$\uparrow$} & \multicolumn{1}{c}{\begin{tabular}[c]{@{}c@{}}PESQ\\ NB$\uparrow$\end{tabular}} & \multicolumn{1}{c}{\begin{tabular}[c]{@{}c@{}}PESQ\\ WB$\uparrow$\end{tabular}} & \multicolumn{1}{c}{UTMOS$\uparrow$} & SIM$\uparrow$ \\ \midrule
w/o FSQ    & 0.915     & \textbf{3.14}       & \textbf{2.52}       & 4.15       &  \textbf{0.83}   \\ \midrule
gvq-32 & 0.912     & 2.91                & 2.30                & 4.06       &  0.74   \\ \midrule
8      & 0.916     & 3.04                & 2.41                & 4.16       &  0.74   \\
16     & 0.919     & 3.08                & 2.45                & 4.15       &  0.77   \\
32     & \textbf{0.922}    & 3.13     & 2.51    & \textbf{4.18}  &  0.80   \\ \bottomrule
\end{tabular}
\label{tab:global_length}
\vspace{-5pt}
\end{table}

\subsection{Control Capabilities of Spark-TTS}
Spark-TTS enables controllable generation by inputting attribute labels or fine-grained attribute values. In label-based control, the model automatically generates the corresponding attribute values (e.g., pitch and speed). However, when these values are manually specified, the system switches to fine-grained control. 

\textbf{Gender} 
To assess Spark-TTS's capability in gender control, we compare it with textual prompt-based controllable TTS models, including VoxInstruct\cite{zhou2024voxinstruct} and Parler-TTS\cite{lyth2024natural}.
For evaluation, we reorganize the test prompts of real speech from PromptTTS~\cite{guo2023prompttts} based on the prompt structures used in VoxInstruct and Parler-TTS. The gender accuracy (Acc) of the generated speech is measured using our gender predictor, which is specifically trained for gender annotation.
The results, presented in Table~\ref{table:gender_control}, show that Spark-TTS significantly outperforms other controllable TTS systems in gender control, demonstrating its strong capability in attribute-based voice generation.

\begin{table}[tp]
\centering
\small
\setlength{\tabcolsep}{3mm}
\caption{Gender control performance of various models.}
\begin{tabular}{lccc}
\toprule
Method      & VoxInstruct    & Parler-tts   & Spark-TTS \\ \midrule
Acc (\%)$\uparrow$    & 82.99          & 98.12     & \textbf{99.77}    \\ \bottomrule
\end{tabular}
\vspace{-10pt}
\label{table:gender_control}
\end{table}

\begin{figure}[tbp]
    \centering
    \vspace{-5pt}
   \subfloat[Pitch for Male]{
        \includegraphics[width=0.45\linewidth]{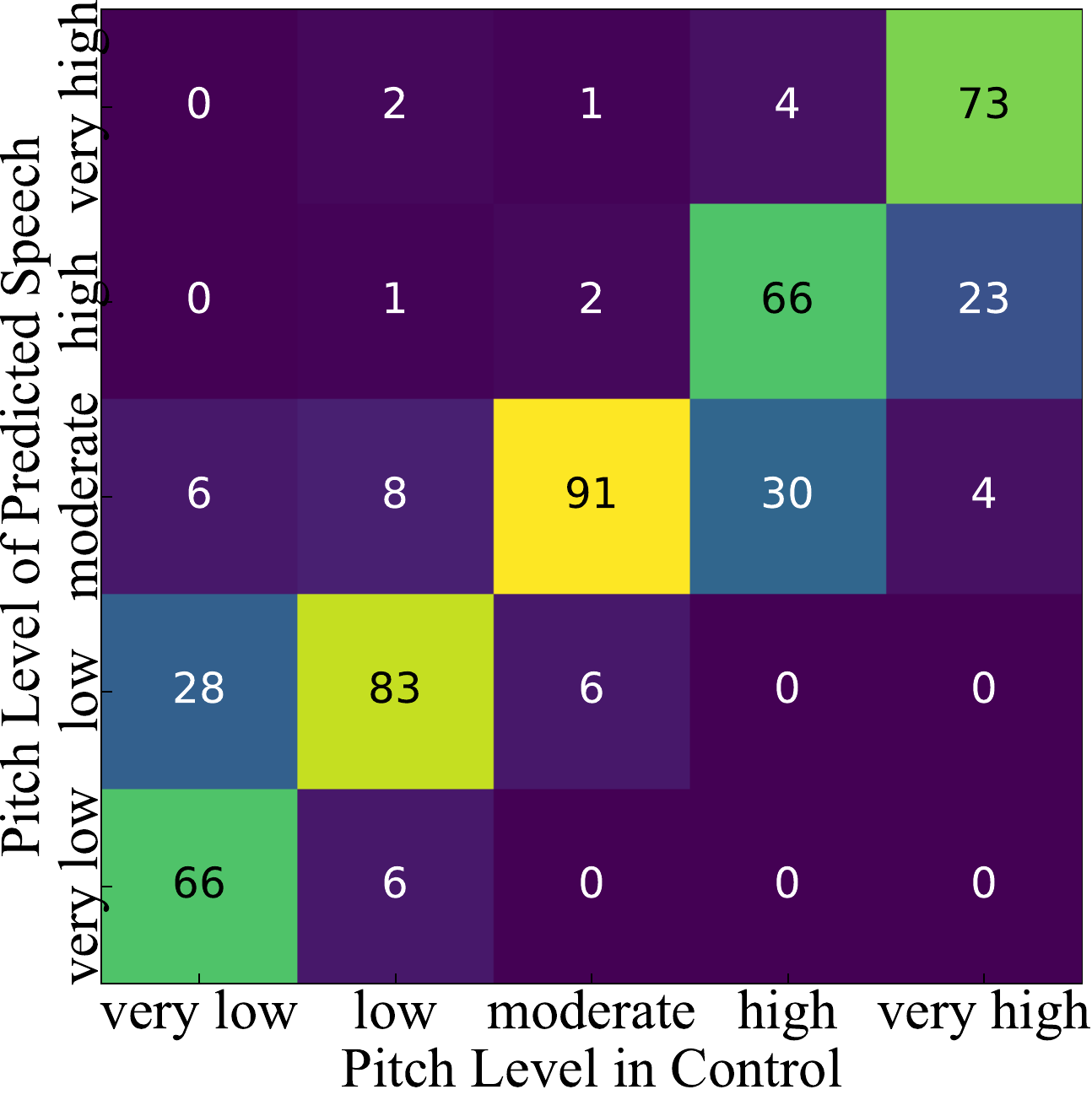}
    }
    \subfloat[Pitch for Female]{
        \includegraphics[width=0.45\linewidth]{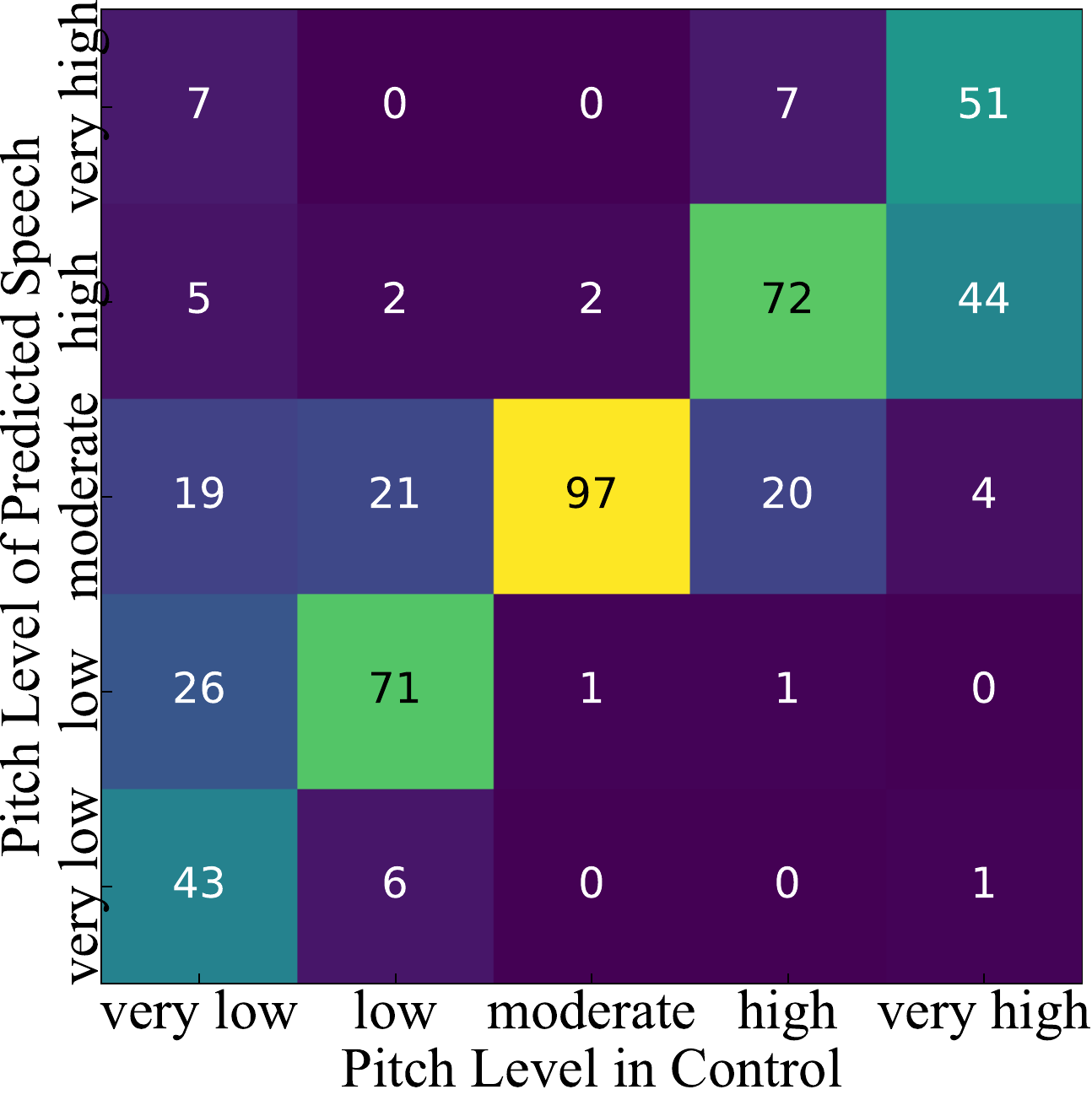}
    } \\
    \subfloat[Speed for English]{
        \includegraphics[width=0.45\linewidth]{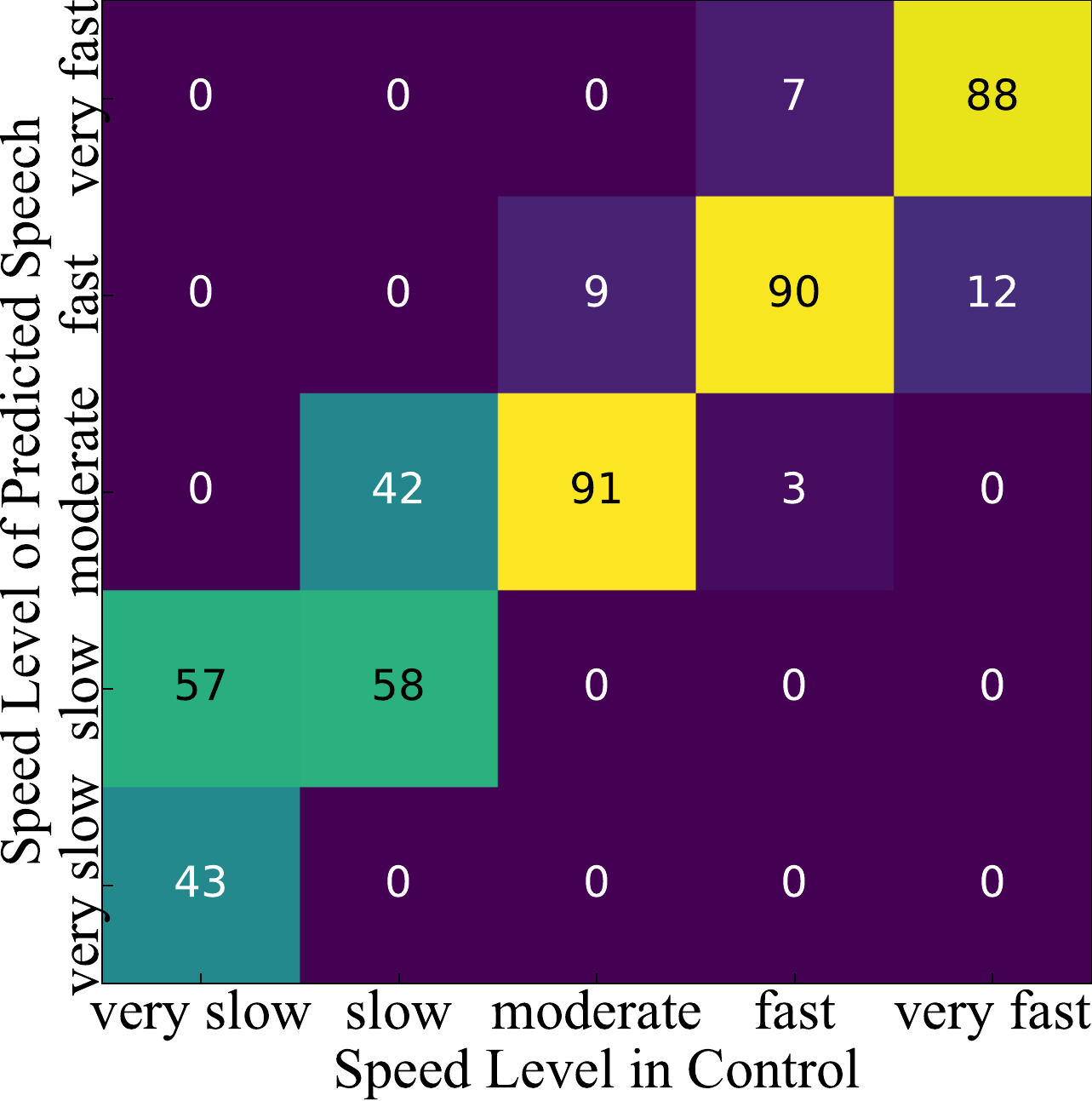}
    }
    \subfloat[Speed for Chinese]{
        \includegraphics[width=0.45\linewidth]{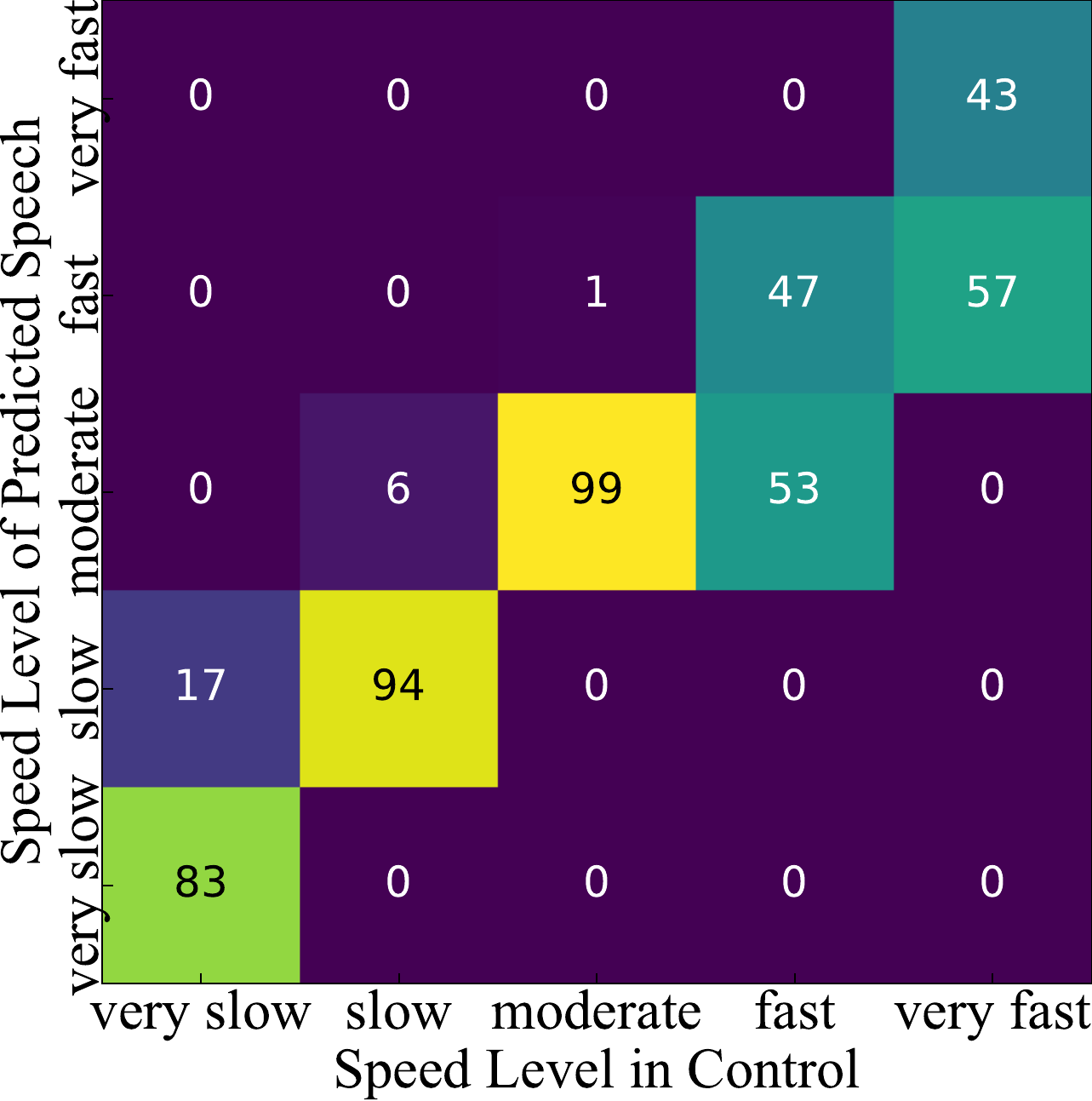}
    } \\
    \caption{Confusion matrix of coarse-grained pitch and speed control results. In pitch-controllable generation, each label's generated samples consist of 50 Chinese and 50 English samples. In speed-controllable generation, each label's generated samples consist of 50 male and 50 female samples. }
    \label{fig:control_matrix}
    \vspace{-10pt}
\end{figure}

\begin{figure}[H]
    \centering
    \vspace{-10pt}
    \subfloat[Pitch for Male]{
        \includegraphics[width=0.47\linewidth]{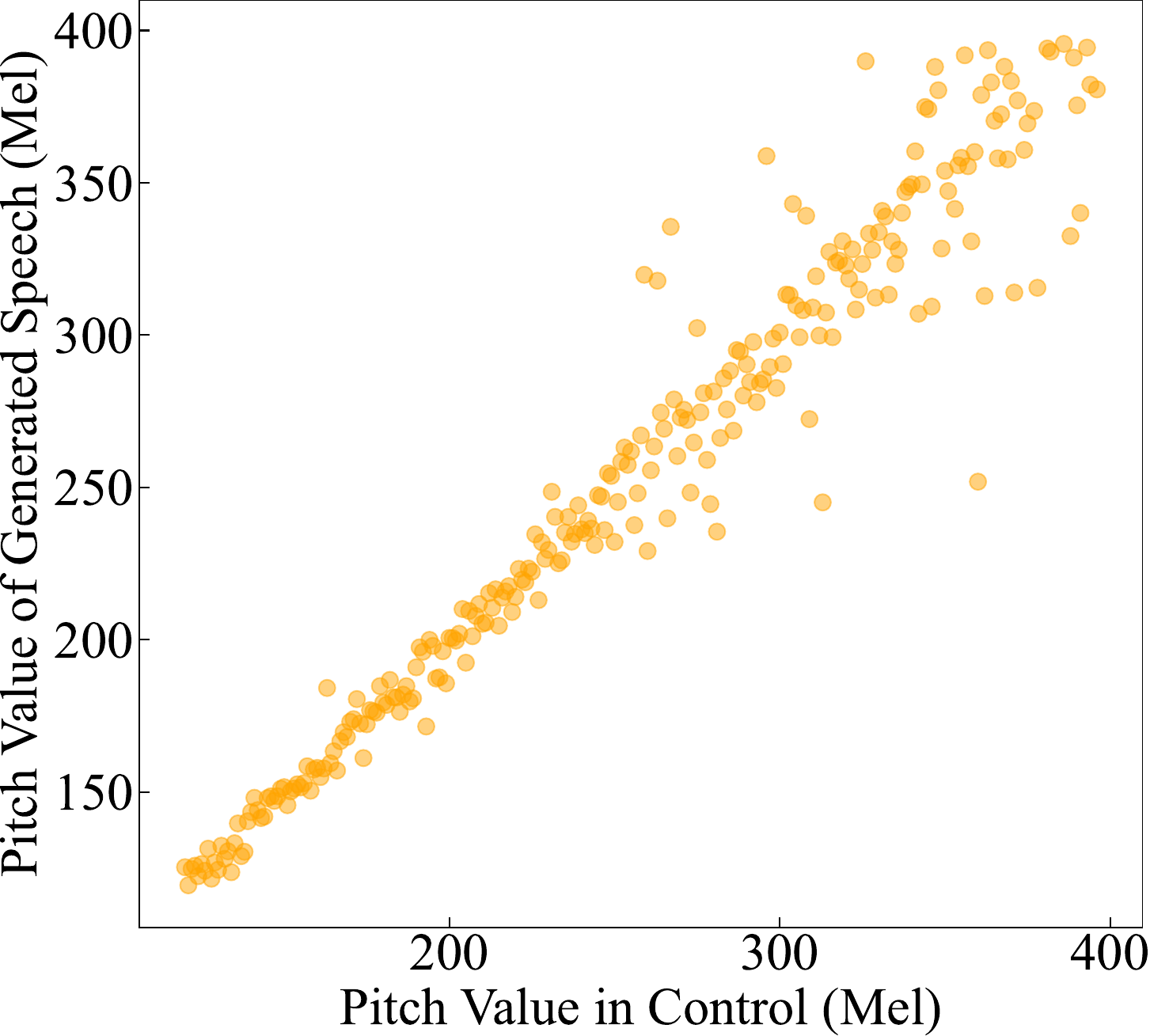}
    }
    \subfloat[Pitch for Female]{
        \includegraphics[width=0.47\linewidth]{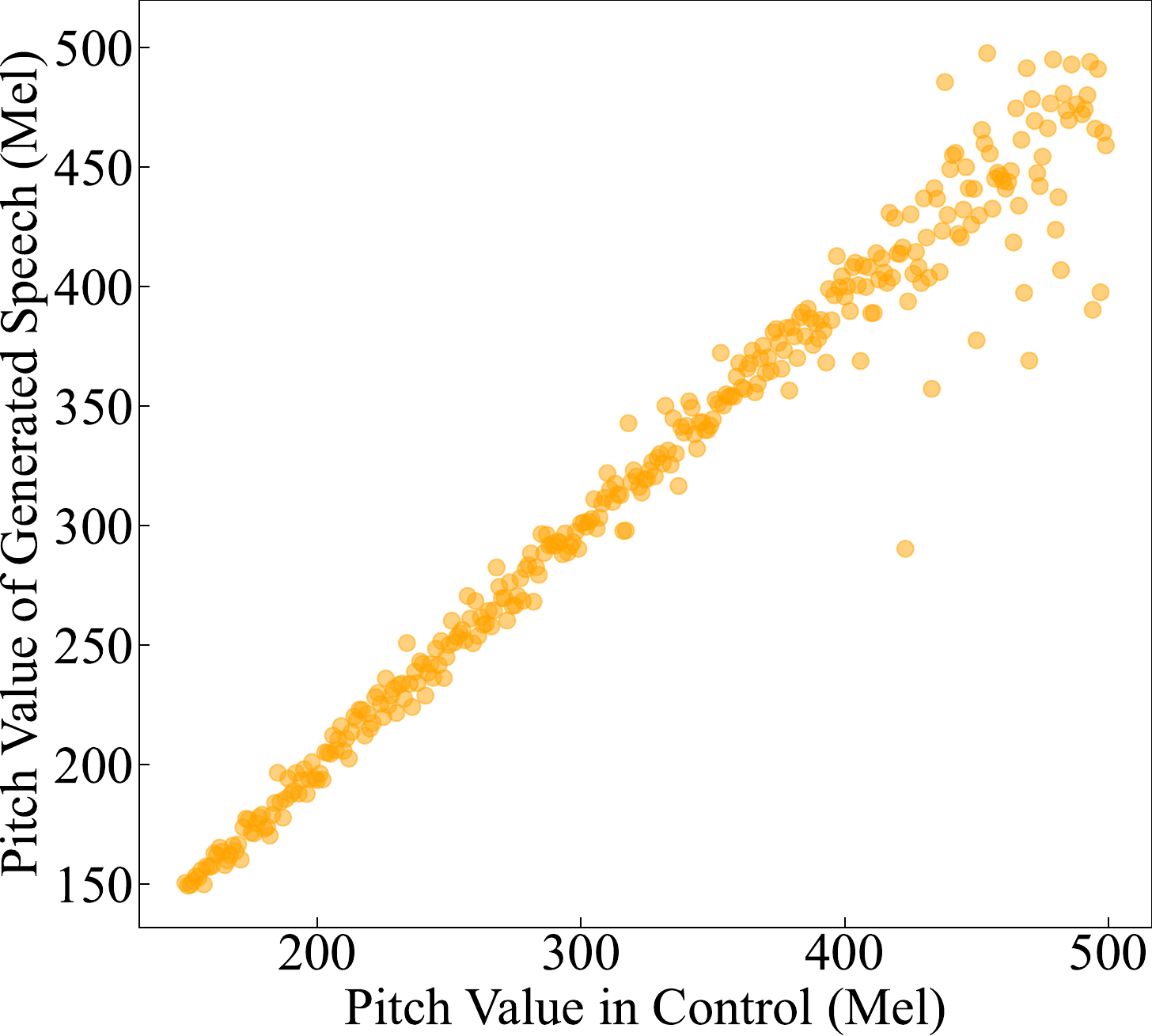}
    } \\
    \subfloat[Speed for English]{
        \includegraphics[width=0.47\linewidth]{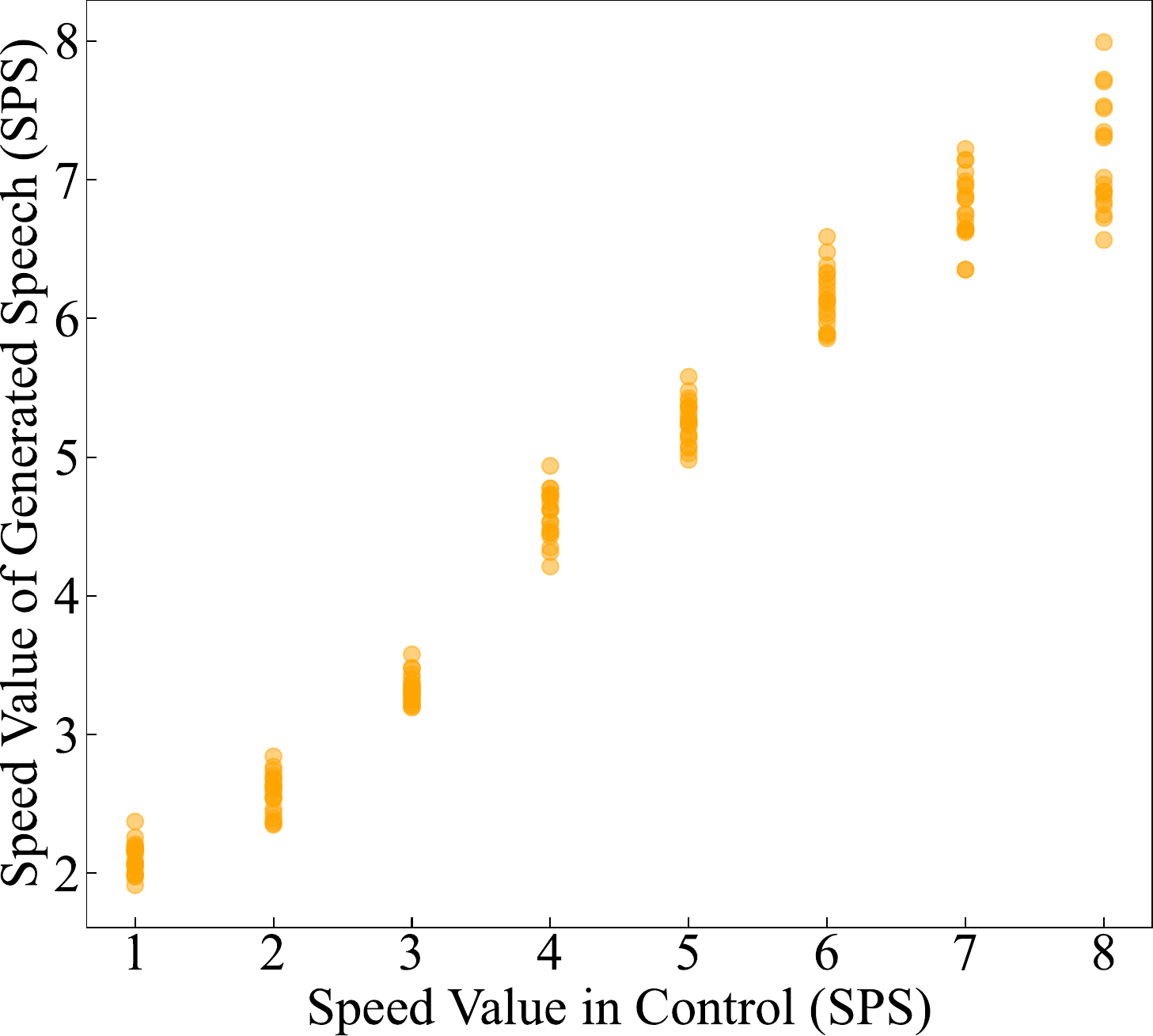}
    }
    \subfloat[Speed for Chinese]{
        \includegraphics[width=0.47\linewidth]{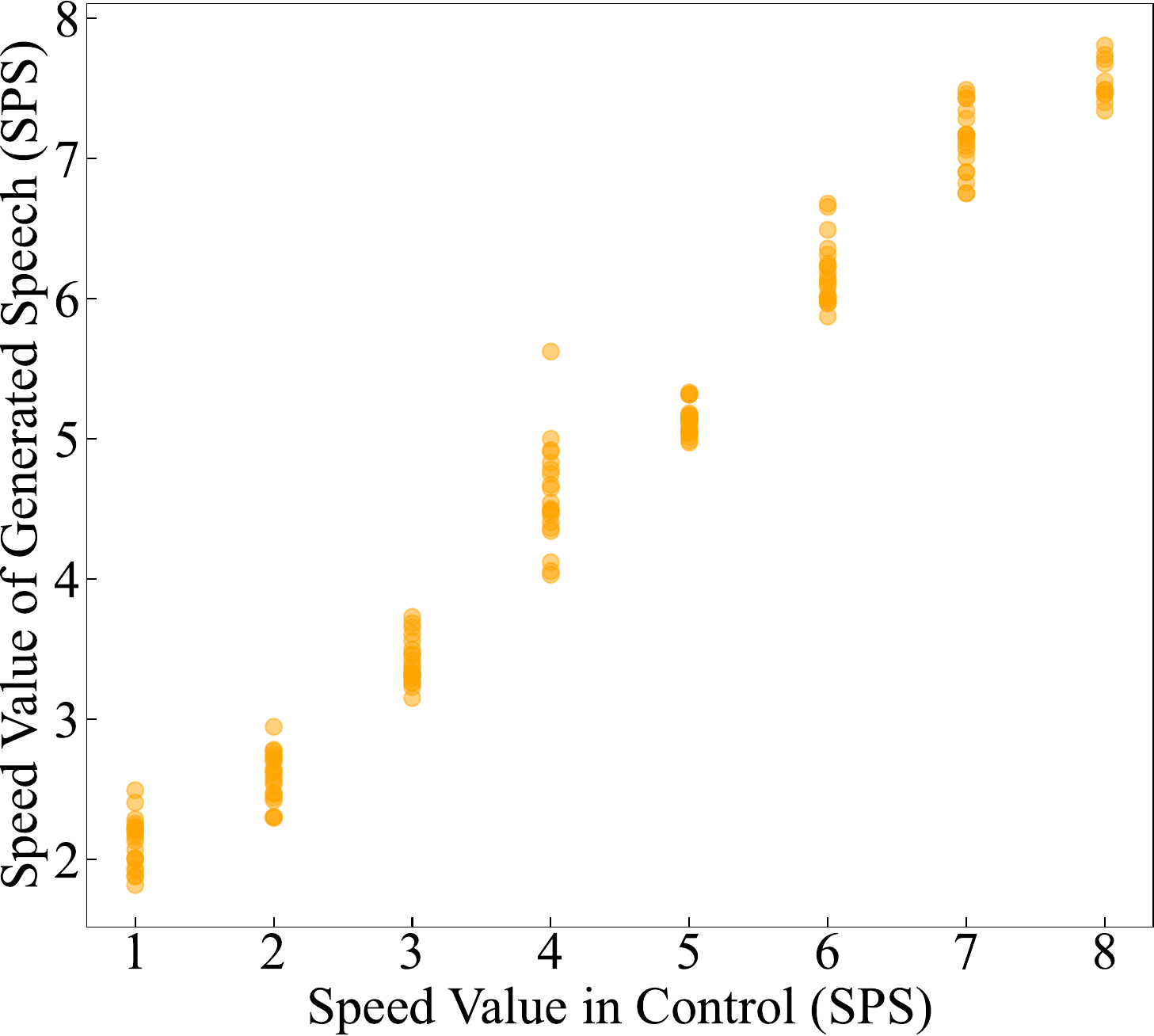}
    } \\
    \caption{Fine-grained pitch and speed control results. For pitch-controllable generation, each generated value includes one Chinese sample and one English sample. For speed-controllable generation, each generated value includes 10 male samples and 10 female samples.}
    \label{fig:control_value}
    \vspace{-15pt}
\end{figure}

\textbf{Pitch and Speed}
Spark-TTS enables controllable generation by inputting attribute labels or fine-grained attribute values. In label-based control, the model automatically generates the corresponding attribute values (e.g., pitch and speed). However, when these values are manually specified, the system switches to fine-grained control. Fig.~\ref{fig:control_matrix} illustrates the control confusion matrices for pitch and speaking rate based on coarse-grained labels, while Fig.~\ref{fig:control_value} presents the fine-grained control performance for pitch and speed. As shown, Spark-TTS accurately generates speech that aligns with the specified attribute labels, demonstrating precise control over both coarse-grained and fine-grained attributes.

\subsection{Zero-shot TTS Performance}
To evaluate Spark-TTS's zero-shot TTS capability, we assess its performance on Seed-TTS-eval and compare it with existing zero-shot TTS models. The results are presented in Table~\ref{tab:zst_tts}, where speech intelligibility is evaluated using the Character Error Rate (CER) for Chinese and the WER for English, following the Seed-TTS-eval\footnote{\scriptsize{\url{https://github.com/BytedanceSpeech/seed-tts-eval}}}. As can been seen, Spark-TTS demonstrates significant superiority in intelligibility for zero-shot TTS scenarios. On test-zh, Spark-TTS achieves a CER second only to the closed-source model Seed-TTS, while it ranks second only to F5-TTS~\cite{chen2024f5} for English WER. This high intelligibility is partly attributed to the semantic feature-based BiCodec and further validates the high quality of our VoxBox dataset in terms of transcripts.
In terms of speaker similarity, while Spark-TTS is relatively weaker than multi-stage or NAR-based methods, it significantly outperforms the single-stage model Llasa~\cite{ye2025llasa}. Notably, Spark-TTS, with just 0.5B model parameters and 100k hours of training data, surpasses Llasa, which has 8B parameters and is trained on 250k hours of data.

Following CosyVoice2~\cite{du2024cosyvoice2}, we evaluate the quality of the generated speech on the LibriSpeech test-clean set. As shown in Table~\ref{tb:tts_utmos}, our method produces audio of significantly higher quality than the original and outperforms CosyVoice2, the SOTA open-source TTS model with multi-stage modeling. This demonstrates the strong performance of Spark-TTS in terms of speech quality.

\begin{table}[htp]
\centering
\small
\caption{Results of Spark-TTS and recent TTS models on the Seed test sets (test-zh for Chinese and test-en for English). † denotes closed-sourced models.}
\setlength{\tabcolsep}{2.2mm}
\begin{tabular}{lcccc}
\toprule
\multicolumn{1}{l|}{\multirow{2}{*}{Model}} & \multicolumn{2}{c|}{test-zh}      & \multicolumn{2}{c}{test-en} \\ \cline{2-5} 
\multicolumn{1}{l|}{}                       & CER$\downarrow$  & \multicolumn{1}{c|}{SIM$\uparrow$}   & WER$\downarrow$          & SIM$\uparrow$          \\ \midrule
\multicolumn{5}{c}{Multi-Stage or NAR Methods}                                                                \\ \midrule
\multicolumn{1}{l|}{Seed-TTS\textsuperscript{†}}  & \textbf{1.12} & \multicolumn{1}{c|}{ \textbf{0.796}} & 2.25         & \textbf{0.762}        \\
\multicolumn{1}{l|}{FireRedTTS}             & 1.51 & \multicolumn{1}{c|}{0.635} & 3.82         & 0.460        \\
\multicolumn{1}{l|}{MaskGCT}                & 2.27 & \multicolumn{1}{c|}{0.774} & 2.62         & 0.714        \\
\multicolumn{1}{l|}{E2 TTS (32 NFE)\textsuperscript{†}}        & 1.97 & \multicolumn{1}{c|}{0.730} & 2.19         & 0.710        \\
\multicolumn{1}{l|}{F5-TTS (32 NFE)}        & 1.56 & \multicolumn{1}{c|}{0.741} & \textbf{1.83}         & 0.647        \\
\multicolumn{1}{l|}{CosyVoice}              & 3.63 & \multicolumn{1}{c|}{0.723} & 4.29         & 0.609        \\
\multicolumn{1}{l|}{CosyVoice2}             & 1.45 & \multicolumn{1}{c|}{0.748} & 2.57         & 0.652        \\ \midrule
\multicolumn{5}{c}{One-Stage AR Methods}                                                                      \\ \midrule
\multicolumn{1}{l|}{Llasa-1B-250k}          & 1.89 & \multicolumn{1}{c|}{0.669} & 3.22         & 0.572        \\
\multicolumn{1}{l|}{Llasa-3B-250k}          & 1.60 & \multicolumn{1}{c|}{0.675} & 3.14         & 0.579        \\
\multicolumn{1}{l|}{Llasa-8B-250k}          & 1.59 & \multicolumn{1}{c|}{\textbf{0.684}} & 2.97         & 0.574        \\ \midrule
\multicolumn{1}{l|}{Spark-TTS}              & \textbf{1.20} & \multicolumn{1}{c|}{0.672} &   \textbf{1.98}       & \textbf{0.584}             \\ \bottomrule
\end{tabular}
\label{tab:zst_tts}
\vspace{-5pt}
\end{table}

\begin{table}[htp]
\small
\vspace{-5pt}
\centering
\caption{Quality comparison of zero-shot TTS audio generation on the LibriSpeech test-clean set. GT represents ground truth.}
\setlength{\tabcolsep}{1.4mm}
\begin{tabular}{lcccc}
\toprule
Method & \multicolumn{1}{c}{GT} & \multicolumn{1}{c}{CosyVoice} & \multicolumn{1}{c}{CosyVoice2} & Spark-TTS \\ \midrule
UTMOS$\uparrow$  & 4.08  &      4.09  &      4.23    & \textbf{4.35}      \\ \bottomrule
\end{tabular}
\label{tb:tts_utmos}
\vspace{-10pt}
\end{table}



\section{Conclusion}
This paper introduces BiCodec, which retains the advantages of semantic tokens, including high compression efficiency and high intelligibility, while addressing the limitation of traditional semantic tokens, which cannot control timbre-related attributes within an LM, by incorporating global tokens. BiCodec achieves a new SOTA reconstruction quality, operating at 50 TPS with a bit rate of 0.65 kbps, surpassing other codecs within the sub-1 kbps range.
Building on BiCodec, we develop Spark-TTS, a text-to-speech model that integrates the textual language model Qwen2.5. Spark-TTS enables voice generation based on specified attributes and supports zero-shot synthesis. To our knowledge, this is the first TTS model to offer fine-grained control over both pitch and speaking rate, while simultaneously supporting zero-shot TTS.
Additionally, to facilitate comparative research, we introduce VoxBox, an open-source dataset designed for controllable speech synthesis. VoxBox not only filters out low-quality textual data but also provides comprehensive annotations, including gender, pitch, and speaking rate, significantly enhancing training for controlled generation tasks.

\clearpage
\section*{Limitation}
Despite its advantages, Spark-TTS also has notable limitations. Similar to Llasa~\cite{ye2025llasa}, which relies on a single codebook and a textual language model, Spark-TTS exhibits relatively lower speaker similarity metrics in zero-shot TTS compared to multi-stage or NAR methods. This may be due to the greater speaker variability introduced by the AR language model during inference.
Currently, Spark-TTS does not impose additional disentanglement constraints between global tokens and semantic tokens. In future work, we aim to enhance global token control over timbre by introducing perturbations to formants or pitch in the semantic token input. This approach will promote better disentanglement of timbre information, allowing BiCodec’s decoder to exert absolute control over timbre. By doing so, we aim to reduce randomness introduced by the AR model, improving the speaker similarity in zero-shot synthesis.


\bibliography{reference}

\newpage

\appendix

\section{BiCodec}
\label{append:bicodec}

\begin{figure*}[htp]
    \centering
    \includegraphics[width=0.95\linewidth]{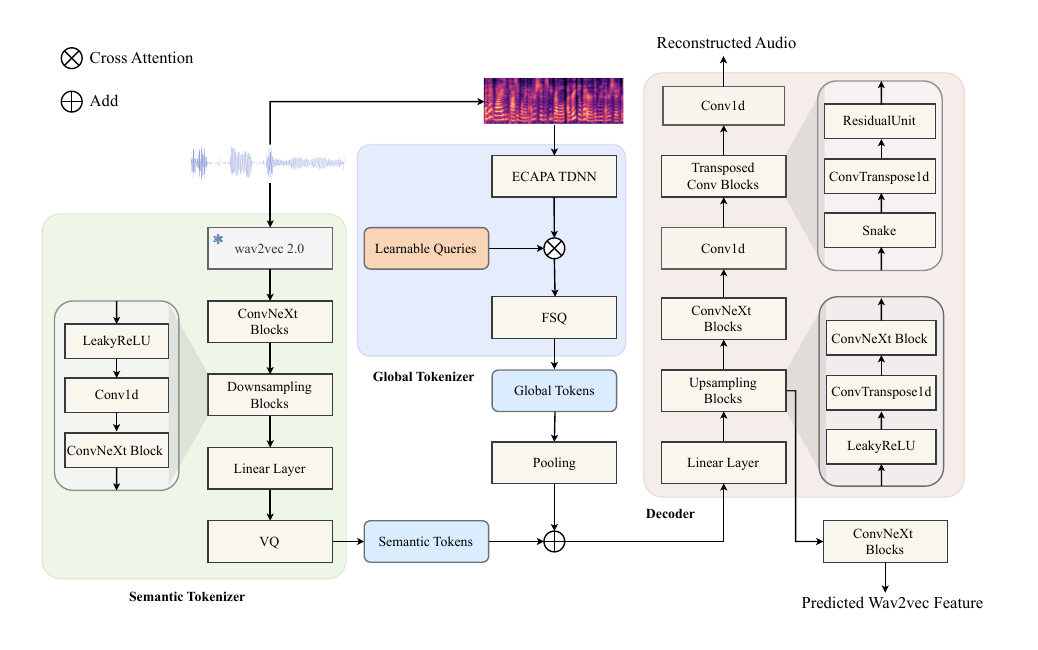}
    \caption{Model Structure of BiCodec}
    \label{fig:bicodec model struture}
    \vspace{-10pt}
\end{figure*}

The model structure of BiCodec is illustrated in Fig.~\ref{fig:bicodec model struture}. BiCodec primarily consists of three components:
\begin{itemize}
    \item Semantic Tokenizer
    \item Global Tokenizer
    \item Decoder
\end{itemize}
Additionally, to compute the feature loss with the input wav2vec 2.0 features, an extra ConvNeXt block is incorporated to predict wav2vec 2.0 features, to further ensure the semantic relevance.

\subsection{Model Configurations}
The semantic tokenizer consists of 12 ConvNeXt blocks and 2 downsampling blocks. The downsampling blacks is only for semantic codes with lower than 50 TPS. The codebook size of VQ is 8192. The ECAPA-TDNN in the global tokenizer features an embedding dimension of 512. Meanwhile, the vector number of the learnable queries in the global tokenizer equal to the final goal token sequence length. For the FSQ module, the FSQ dimension is set to 6, with each dimension having 4 levels, resulting in a codebook size of 4096.

The upsampling rates in the Transposed Convolution Blocks are set to [8, 5, 4, 2] for 16 kHz sampled audio and [8, 5, 4, 3] for 24 kHz sampled audio. The reconstruction performance of BiCodec with 24 kHz sampled audio is presented in Table~\ref{tab:codec_other_data}.

\subsection{Compared Methods}
\label{sc:codec_compare_append}

\begin{itemize}
    \item \textbf{Encodec} \cite{defossez2022high}: An RVQ-based codec designed for universal audio compression.
    
    \item \textbf{DAC} \cite{kumar2024high}: An RVQ-based codec for universal audio.
    
    \item \textbf{Mimi} \cite{defossez2024moshi}: An RVQ-based codec with semantic constraint for speech.
    
    \item \textbf{Single-Codec} \cite{li2024single}: A single-stream Mel codec that incorporates speaker embeddings. The reconstruction results for this method are provided by the authors.

    \item \textbf{BigCodec} \cite{xin2024bigcodec}: A VQ-based single-stream codec for speech.
    
    \item \textbf{SpeechTokenizer} \cite{zhang2023speechtokenizer}: An RVQ-based codec with semantic distillation for speech.
    
    \item \textbf{X-codec} \cite{ye2024codec}: An RVQ-based codec with semantic distillation for speech.

    \item \textbf{X-codec2} \cite{ye2025llasa}: A FSQ-based single-stream codec with semantic distillation for speech.
    
    \item \textbf{StableCodec} \cite{parker2024scaling}: A residual FSQ-based tokenizer for speech.
   
    \item \textbf{WavTokenizer} \cite{ji2024wavtokenizer}: A single VQ codebook-based tokenizer for universal audio.
    
\end{itemize}

\subsection{Additional Experiment}
\label{sc:codec_exp_append}

To evaluate the performance of BiCodec at lower bitrates, we apply a downsampling operation in the semantic encoder, reducing the semantic token rate to 25 TPS. We compare BiCodec with Single-Codec~\cite{li2024single}, which operates at a similar bitrate, on the LibriSpeech test-clean and LibriTTS test-clean datasets. The results are presented in Table~\ref{tab:lower_bit} and Table~\ref{tab:lower_bit_libritts}.

\begin{table*}[htb]
\centering
\caption{Performance of BiCodec with lower bitrate on the LibriSpeech test-clean dataset.}
\setlength{\tabcolsep}{1.mm}
\begin{tabular}{lccccccccc}
\toprule
Model           & \begin{tabular}[c]{@{}c@{}}Codebook \\ Size\end{tabular} & Nq & \begin{tabular}[c]{@{}c@{}}Token \\ Rate\end{tabular} & Bandwidth & STOI$\uparrow$ & \begin{tabular}[c]{@{}c@{}}PESQ\\ NB$\uparrow$\end{tabular} & \begin{tabular}[c]{@{}c@{}}PESQ\\ WB$\uparrow$\end{tabular} & UTMOS$\uparrow$ & SIM$\uparrow$  \\ \midrule
Single-Codec     & 8192    & 1  & 23.4     & 304   & 0.86    & 2.42      & 1.88           & 3.72  & 0.60  \\ \midrule
BiCodec-4096-25  & 4096    & 1  & 25      &  300   & 0.88    & 2.53      & 1.97           & 4.00  & 0.70   \\
BiCodec-8192-25  & 8192    & 1  & 25      &  325   & 0.89    & 2.62      & 2.05           & 4.13  & 0.71    \\
BiCodec-4096-50  & 4096    & 1  & 50      &  600   & 0.92    & 3.03      & 2.42           & 4.17  & 0.78    \\
\bottomrule
\end{tabular}
\label{tab:lower_bit}
\end{table*}

\begin{table*}[]
\centering
\caption{Reconstruction performance of BiCodec with various bitrates on the LibriTTS test-clean dataset.}
\setlength{\tabcolsep}{1.6mm}
\begin{tabular}{lcccccccl}
\hline
\begin{tabular}[c]{@{}c@{}}Codebook \\ Size\end{tabular}  & Nq & \begin{tabular}[c]{@{}c@{}}Token \\ Rate\end{tabular} & Bandwidth & STOI$\uparrow$          & \begin{tabular}[c]{@{}c@{}}PESQ\\ NB$\uparrow$\end{tabular} & \begin{tabular}[c]{@{}c@{}}PESQ\\ WB$\uparrow$\end{tabular} & UTMOS$\uparrow$           & SIM$\uparrow$ \\ \hline
4096          & 1  & 25     & 300    & 0.88     & 2.47      & 1.91     & 3.88   & 0.67    \\
8192          & 1  & 25     & 325    & 0.88     & 2.56      & 1.98     & 4.02   & 0.68    \\
4096          & 1  & 50     & 600    & 0.91     & 2.96      & 2.36     & 4.10   & 0.75    \\
8192          & 1  & 50     & 650    & \textbf{0.92} & \textbf{3.08}   & \textbf{2.46}  & \textbf{4.11} & \textbf{0.78}    \\ \hline
\end{tabular}
\label{tab:lower_bit_libritts}
\end{table*}

\textbf{Global Token Length} The reconstruction performance of BiCodec with varying global token lengths on the LibriTTS test-clean dataset is presented in Table~\ref{tab:glength_libritts}.

\begin{table}[]
\centering
\caption{Performance of BiCodec with varying global token lengths for reconstruction on the LibriTTS test-clean dataset, where "w/o" indicates the omission of FSQ-based quantization, and gvq-32 means the global tokenizer is implemented with group VQ.}
\setlength{\tabcolsep}{1.mm}
\begin{tabular}{lccccc}
\hline
\multicolumn{1}{c}{\begin{tabular}[c]{@{}c@{}}Global\\ Token\end{tabular}} & \multicolumn{1}{c}{STOI$\uparrow$ } & \multicolumn{1}{c}{\begin{tabular}[c]{@{}c@{}}PESQ\\ NB$\uparrow$ \end{tabular}} & \multicolumn{1}{c}{\begin{tabular}[c]{@{}c@{}}PESQ\\ WB$\uparrow$ \end{tabular}} & \multicolumn{1}{c}{UTMOS$\uparrow$ } & \multicolumn{1}{l}{SIM$\uparrow$ } \\ \hline
w/o      & \textbf{0.923}    & \textbf{3.1}    & \textbf{2.48}    & 4.09            &  \textbf{0.81}          \\ \hline
gvq-32   & 0.913             & 2.91            & 2.30              & 4.06           &  0.71          \\ \hline
8        & 0.916             & 2.97            & 2.34             & 4.10             &  0.72          \\
16       & 0.918             & 3.03            & 2.40              & 4.08            &  0.74          \\
32       & 0.921    & \textbf{3.08}   & \textbf{2.46}    & \textbf{4.11}   &  0.78          \\ \hline
\end{tabular}
\label{tab:glength_libritts}
\end{table}

\textbf{Performance on Other Datasets} 
To evaluate the generalization ability of BiCodec, we conducted experiments on a broader range of diverse datasets. The results are presented in Table~\ref{tab:codec_other_data}.

\begin{table*}[htp]
\centering
\setlength{\tabcolsep}{2.7mm}
\caption{Reconstruction performance on various datasets: Data-P comprises low-quality Chinese recordings made by internal staff using mobile phones; Data-S consists of expressive Chinese data recorded in a professional studio; and Data-M is a multilingual dataset collected from in-the-wild sources.}
\begin{tabular}{lcccccccc}
\hline
Data                    & Method      & \begin{tabular}[c]{@{}c@{}}Codebook\\ Size\end{tabular} & \begin{tabular}[c]{@{}c@{}}Traing\\ Data\end{tabular} & STOI$\uparrow$          & \begin{tabular}[c]{@{}c@{}}PESQ\\ NB$\uparrow$\end{tabular} & \begin{tabular}[c]{@{}c@{}}PESQ\\ WB$\uparrow$\end{tabular} & UTMOS$\uparrow$         & SIM$\uparrow$           \\ \hline
\multirow{3}{*}{Data-P} & X-codec2    & 65536       & 150k      & 0.89     & 2.69        & 2.10         & 3.16      & 0.73          \\
                        & BiCodec     & 8192        & 3k        & \textbf{0.90}  & \textbf{2.80}     & \textbf{2.22}     & \textbf{3.22}    & \textbf{0.78} \\
                        & BiCodec-24k & 8192        & 20k & \textbf{0.90} & \textbf{2.80}     & 2.19     & 3.20      & \textbf{0.78}  \\ \hline
\multirow{3}{*}{Data-S} & X-codec2    & 65536       & 150k      & 0.92     & 2.81       & 2.30     & 3.16     & 0.69        \\
                        & BiCodec     & 8192        & 3k        & 0.93          & \textbf{3.04}    & \textbf{2.50}      & \textbf{3.28}   & \textbf{0.82} \\
                        & BiCodec-24k & 8192        & 20k       & 0.93          & 3.00             & 2.44       & 3.24       & \textbf{0.82} \\ \hline
\multirow{3}{*}{Data-M} & X-codec2    & 65536       & 150k      & 0.84          & 2.43             & 1.87       & 2.17       & 0.75          \\
                        & BiCodec     & 8192        & 3k        & \textbf{0.85} & 2.56             & \textbf{1.91}       & 2.17          & \textbf{0.76} \\
                        & BiCodec-24k & 8192        & 20k       & \textbf{0.85} & \textbf{2.57}    & \textbf{1.91}       & \textbf{2.28} & \textbf{0.76} \\ \hline
\end{tabular}
\label{tab:codec_other_data}
\end{table*}

\section{Inference of Spark-TTS}
\textbf{Zero-shot TTS} There are two inference strategies for zero-shot TTS:
\begin{itemize}
    \item Using the text to be synthesized along with the global tokens from a reference audio as the prompt to generate speech, e.g.,
    [<\textit{content text}> <\textit{global token}> $\rightarrow$ <\textit{semantic token}>].

    \item Incorporating both the transcript and semantic tokens of the reference audio as a prefix in the prompt, e.g., [<\textit{content text}> <\textit{reference text}> <\textit{global token}> <\textit{semantic token of reference}> $\rightarrow$ <\textit{semantic token}>].
    
    \end{itemize}
Among these, the second approach achieves higher speaker similarity. The results reported in Table~\ref{tab:zst_tts} are based on this second inference strategy. A comparison between the two inference methods is provided in Table~\ref{tab:zst_tts_prefix}.

\textbf{Voice Creation} Controllable TTS includes two levels of control for inference:
\begin{itemize}
    \item Coarse-grained control: The prompt consists of the text to be synthesized along with attribute labels, e.g., [<\textit{content text}> <\textit{attribute label}> $\rightarrow$ <\textit{attribute values}> <\textit{global tokens}> <\textit{semantic token}>]. In this process, the fine-grained attribute values are predicted first, followed by the generation of global tokens and then semantic tokens, in a CoT manner.

    \item Fine-grained control: The prompt includes the text to be synthesized, attribute levels, and precise attribute values, e.g., [<\textit{content text}> <\textit{attribute label}> <\textit{attribute values}> $\rightarrow$ <\textit{global tokens}> <\textit{semantic token}>].
    
\end{itemize}

\section{Compared Zero-shot Methods}
\begin{itemize}
    \item \textbf{Seed-TTS}~\cite{anastassiou2024seed}: A two-stage model that employs an AR LM for semantic token prediction and flow matching for acoustic feature generation.
    
    \item \textbf{FireRedTTS}~\cite{guo2024fireredtts}: A two-stage model similar to Seed-TTS, using an AR LM for semantic tokens and flow matching for acoustic features.
    
    \item \textbf{MaskGCT}~\cite{wang2024maskgct}: A NAR model that applies masking-based generative strategies for speech synthesis.
    
    \item \textbf{E2 TTS }: A flow matching-based model that predicts Mel spectrograms as acoustic features.
    
    \item \textbf{F5-TTS}~\cite{chen2024f5}: A flow matching-based method that also uses Mel spectrograms as acoustic features.
    
    \item \textbf{CosyVoice}~\cite{du2024cosyvoice}:  A two-stage model with an AR LM for semantic token prediction and flow matching for acoustic feature generation.
     
    \item \textbf{CosyVoice2 }~\cite{du2024cosyvoice2}: An improved version of CosyVoice, maintaining the two-stage structure with an AR LM for semantic tokens and flow matching for acoustic features.
    
    \item \textbf{Llasa}~\cite{ye2025llasa}: A single-stream codec-based TTS model that uses a single AR language model for direct single-stream code prediction.
    
\end{itemize}

\begin{table}[htp]
\centering
\caption{Zero-shot performance of Spark-TTS with and without reference audio as a prefix.}
\setlength{\tabcolsep}{0.5mm}
\begin{tabular}{lcccc}
\toprule
\multicolumn{1}{l|}{\multirow{2}{*}{Model}} & \multicolumn{2}{c|}{test-zh}      & \multicolumn{2}{c}{test-en} \\ \cline{2-5} 
\multicolumn{1}{l|}{}                          & CER$\downarrow$  & \multicolumn{1}{c|}{SIM$\uparrow$}   & WER$\downarrow$          & SIM$\uparrow$          \\ \midrule
\multicolumn{1}{l|}{Spark-TTS}                 & 1.20     & \multicolumn{1}{c|}{0.678}      &     1.98        &    0.584          \\ 
\multicolumn{1}{l|}{Spark-TTS w/o prefix}      & \textbf{0.98}     & \multicolumn{1}{c|}{0.628}      &    \textbf{1.32}     &   0.474           \\ \bottomrule
\end{tabular}
\label{tab:zst_tts_prefix}
\vspace{-15pt}
\end{table}

\section{Objective  Metircs}

\begin{itemize}
    \item \textbf{STOI} \cite{andersen2017non}: A widely used metric for assessing speech intelligibility. Scores range from 0 to 1, with higher values indicating better intelligibility.
    
    \item \textbf{PESQ} \cite{rix2001perceptual}:  A speech quality assessment metric that compares the reconstructed speech to a reference speech signal. We evaluate using both wide-band (WB) and narrow-band (NB) settings.
    
    \item \textbf{UTMOS} \cite{saeki2022utmos}: An automatic Mean Opinion Score (MOS) predictor, providing an estimate of overall speech quality.

    \item \textbf{SIM}: A speaker similarity metric, computed as the cosine similarity between the speaker embeddings of the reconstructed speech (generated speech in TTS) and the original input speech (prompt speech in TTS). We extract speaker embeddings using WavLM-large, fine-tuned on the speaker verification task~\cite{chen2022large}.
\end{itemize}

\section{VoxBox}
\label{ap.sec:voxbox}
\subsection{Criteria for Pitch and Speed Categorization}
\label{ap.sec.pitch_class}

\begin{itemize}
    \item \textbf{Speed} The adoption of the 5th, 20th, 80th, and 95th percentiles to segment speech rates into distinct categories is founded on the need to accurately reflect the natural distribution of speech tempo variations within the population. These percentiles help to capture the extremes and the more central values of speech rate, ensuring that each category is meaningful and representative of specific vocal characteristics. 

    \item \textbf{Pitch} Similar to the segmentation of speech rate, the division of pitch also starts from human subjective perception and the actual distribution characteristics. However, because humans are more sensitive to higher frequencies within the range of human fundamental frequencies, the 5th, 20th, 70th, and 90th percentiles are used as the division boundaries.

\end{itemize}

\begin{tcolorbox}[colback=gray!10!white, colframe=gray!80!black, title=Pitch Group for Male, sharp corners]
\begin{tabular}{@{}ll@{}}
Very Low:  & $< 145 $ Mel \\
Low:       & $145 \text{--} 164$ Mel \\
Moderate:  & $164 \text{--} 211$ Mel \\
High:      & $211 \text{--} 250$ Mel \\
Very High: & $>= 250$ Mel \\
\end{tabular}
\end{tcolorbox}

\begin{tcolorbox}[colback=gray!10!white, colframe=gray!80!black, title=Pitch Group for Female, sharp corners]
\begin{tabular}{@{}ll@{}}
Very Low:  & $< 225 $ Mel \\
Low:       & $225 \text{--} 258$ Mel \\
Moderate:  & $258 \text{--} 314$ Mel \\
High:      & $314 \text{--} 353$ Mel \\
Very High: & $>= 353$ Mel \\
\end{tabular}
\end{tcolorbox}

\begin{tcolorbox}[colback=gray!10!white, colframe=gray!80!black, title=Speaking Rate Group for Chinese, sharp corners]
\begin{tabular}{@{}ll@{}}
Very Slow:  & $ < 2.7 $ SPS          \\
Slow:       & $ 2.7 \text{--} 3.6 $  SPS  \\
Moderate:   & $ 3.6 \text{--} 5.2 $ SPS   \\
Fast:       & $ 5.2 \text{--} 6.1 $ SPS   \\
Very Fast:  & $ >= 6.1  $  SPS        \\
\end{tabular}    
\end{tcolorbox}

\begin{tcolorbox}[colback=gray!10!white, colframe=gray!80!black, title=Speaking Rate Group for English, sharp corners]
\begin{tabular}{@{}ll@{}}
Very Slow:  & $ < 2.6 $ SPS \\
Slow:       & $ 2.6 \text{--} 3.4 $  SPS \\
Moderate:   & $ 3.4 \text{--} 4.8 $ SPS \\
Fast:       & $ 4.8 \text{--} 5.5 $ SPS \\
Very Fast:  & $  >= 5.5  $  SPS        \\
\end{tabular}    
\end{tcolorbox}

\subsection{Data for Gender Predictor Training}
\label{sc:gender_data_ap}
 We fine-tune the WavLM-large model for gender classification using datasets that contain explicit gender labels, including VCTK~\cite{yamagishi2019vctk}, AISHELL-3~\cite{shi2020aishell}, MLS-English~\cite{Pratap2020MLSAL}, MAGICDATA~\cite{MAGICDATA2019}, and 
CommonVoice~\cite{ardila2019common}.

\subsection{Annotation}
In addition to the attributes involved in the experiments of this paper, to make VoxBox applicable to a wider range of scenarios, we have also annotated more information for each sample of VoxBox, including age and emotion. Similar to the gender annotations, we fine-tune the WavLM-large model based on AISHELL-3, VCTK, MAGICDATA, CommonVoice, and HQ-Conversations to predict five age ranges: Child, Teenager, Young Adult, Middle-aged, and Elderly. The performance metrics for both the gender and age predictors are presented in Table~\ref{tb:predictor}, where both Wav2vec 2.0-ft~\cite{burkhardt2023speech} and SpeechCraft~\cite{jin2024speechcraft} are based on the pre-trained Wav2vec 2.0 model.

\begin{table}[htp]
\centering
\setlength{\tabcolsep}{3mm}
\caption{Comparison of different models on attribute predictions: All evaluations are conducted on the AISHELL-3 test dataset.}
\begin{tabular}{lcc}
\toprule
Model                                     & Age Acc$\uparrow$       & Gender Acc$\uparrow$    \\ \midrule
wav2vec 2.0-ft & 80.2          & 98.8          \\
SpeechCraft     & 87.7          & 97.7          \\ \midrule
Our                                       & \textbf{95.6} & \textbf{99.4} \\ \bottomrule
\end{tabular}
\label{tb:predictor}
\end{table}

For datasets without emotion labels in the original metadata, we assign various emotion labels, sourced from different models, to the relevant samples. Specifically, we provide the following tags:

\begin{itemize}
    \item \textbf{emotion2vec Emotion}: Emotion label predicted with Emtion2vec~\cite{ma2023emotion2vec}.
    
    \item  \textbf{Confidence Score}: Confidence score of the the predicted emotion2vec label given by emotion2vec.
    
    \item \textbf{SenseVoiceSmall Emotion}: Emotion label predicted with SenseVoiceSmall\footnote{\url{https://huggingface.co/FunAudioLLM/SenseVoiceSmall}}.

    \item \textbf{Text Emotion}: Emotion label predicted with Qwen2.5-72B-Instruct~\footnote{\url{https://huggingface.co/Qwen/Qwen2.5-72B-Instruct}} with text as input. The prompt case for English text can be found in Box
    
\end{itemize}

\begin{tcolorbox}[title=Prompt for Text Emotion Tag (English), colback=white, colframe=gray!80!black, sharp corners]
Please assess the emotion of the following text and select the most appropriate label from these options:

[Fearful, Happy, Disgusted, Sad, Surprised, Angry, Neutral]. 

Please note, only provide the label without any additional description or reasoning. Here is the text: "Clearly, the need for a personal loan is written in the stars."
\label{box:prompt}
\end{tcolorbox}

\subsection{Data Statistics}
\begin{figure*}[h!]
\centering
\begin{subfigure}{.33\textwidth}
  \centering
  \includegraphics[width=\linewidth]{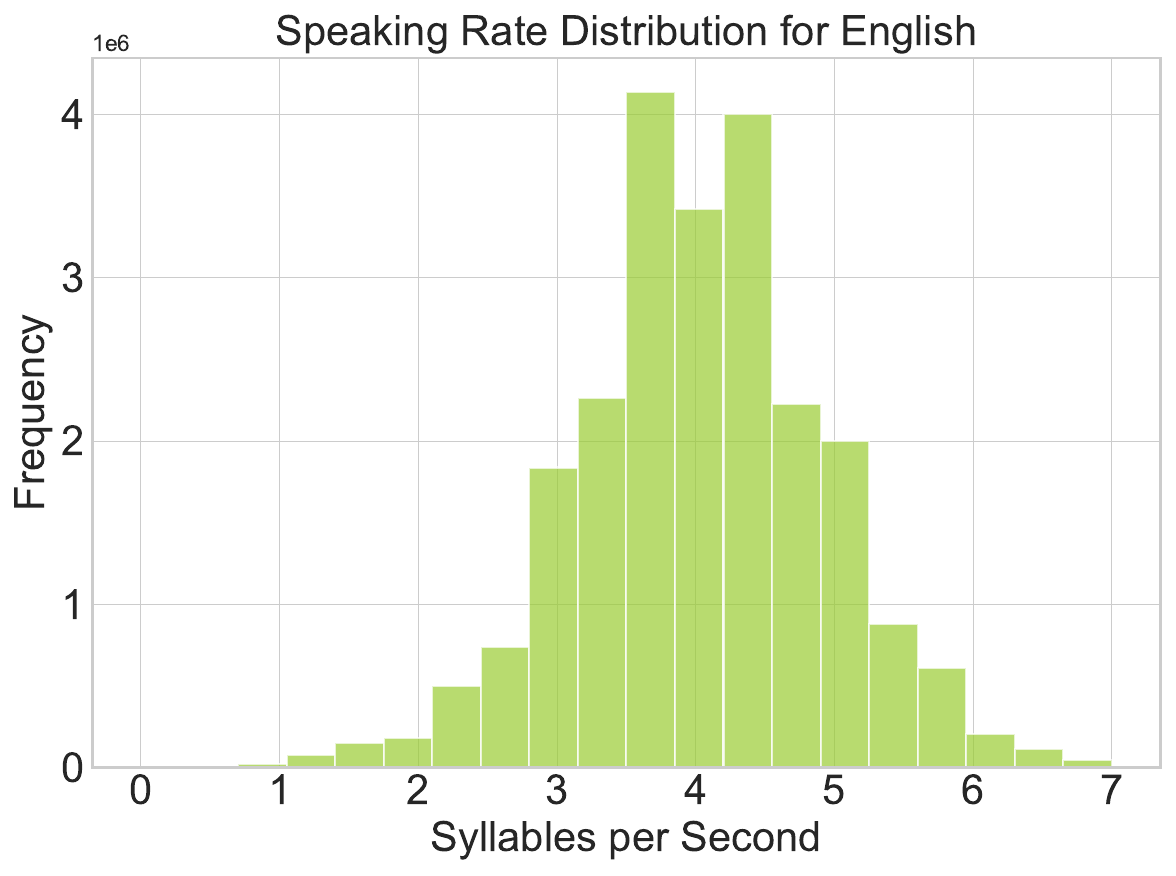}
\end{subfigure}%
\begin{subfigure}{.33\textwidth}
  \centering
  \includegraphics[width=\linewidth]{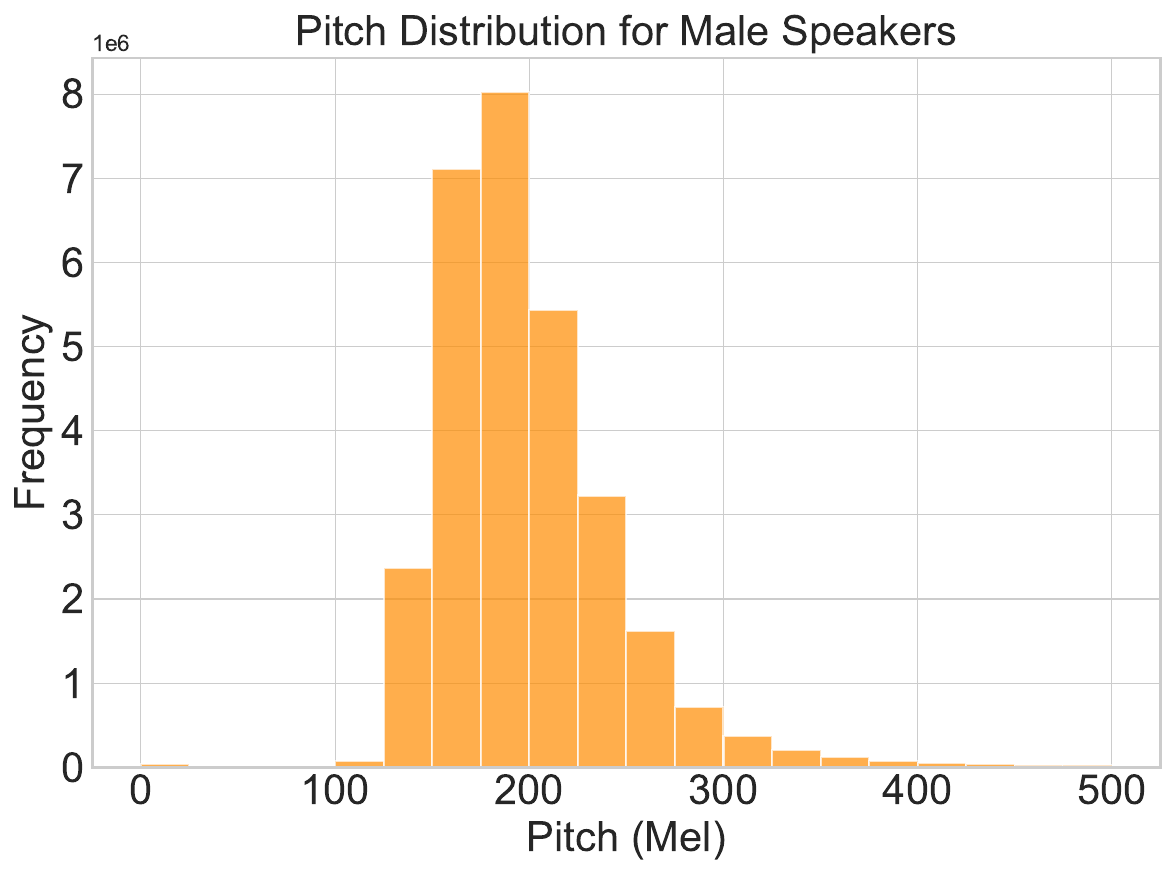}
\end{subfigure}%
\begin{subfigure}{.33\textwidth}
  \centering
  \includegraphics[width=\linewidth]{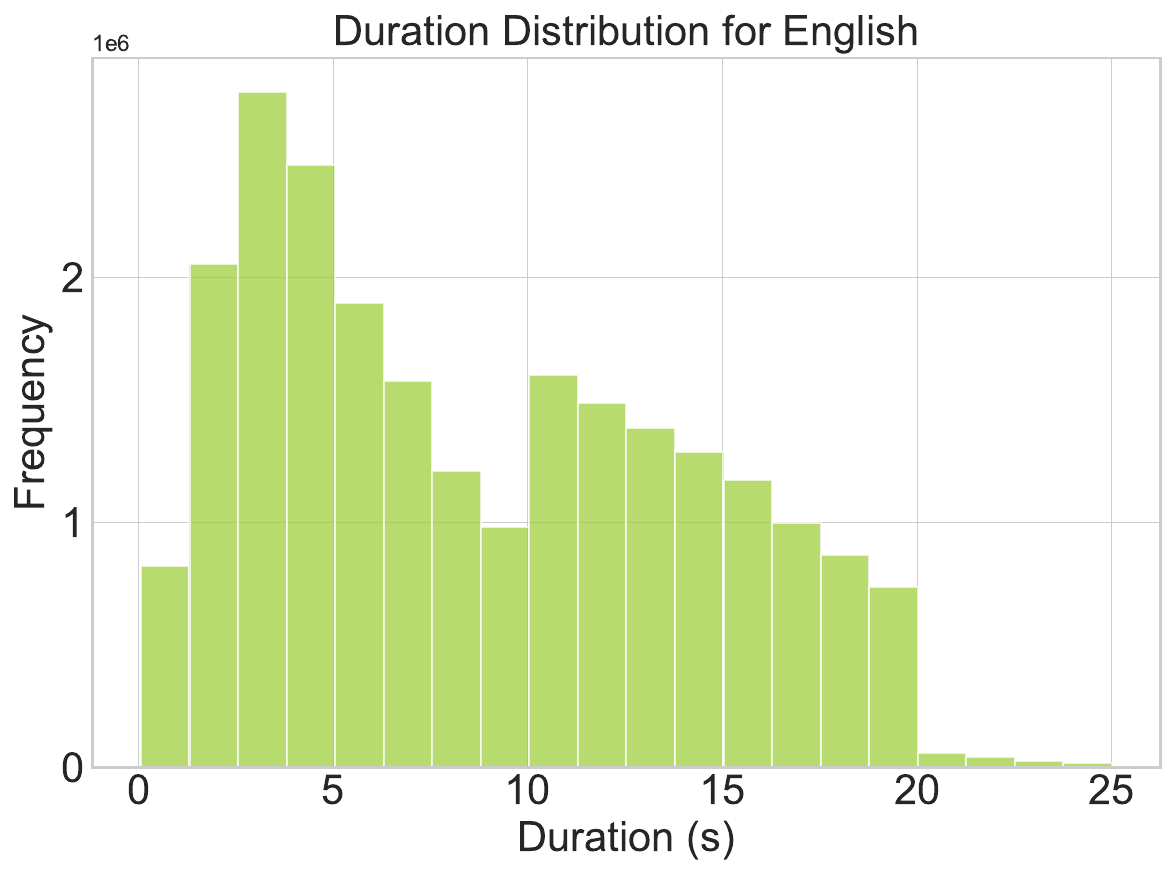}
\end{subfigure}
\begin{subfigure}{.33\textwidth}
  \centering
  \includegraphics[width=\linewidth]{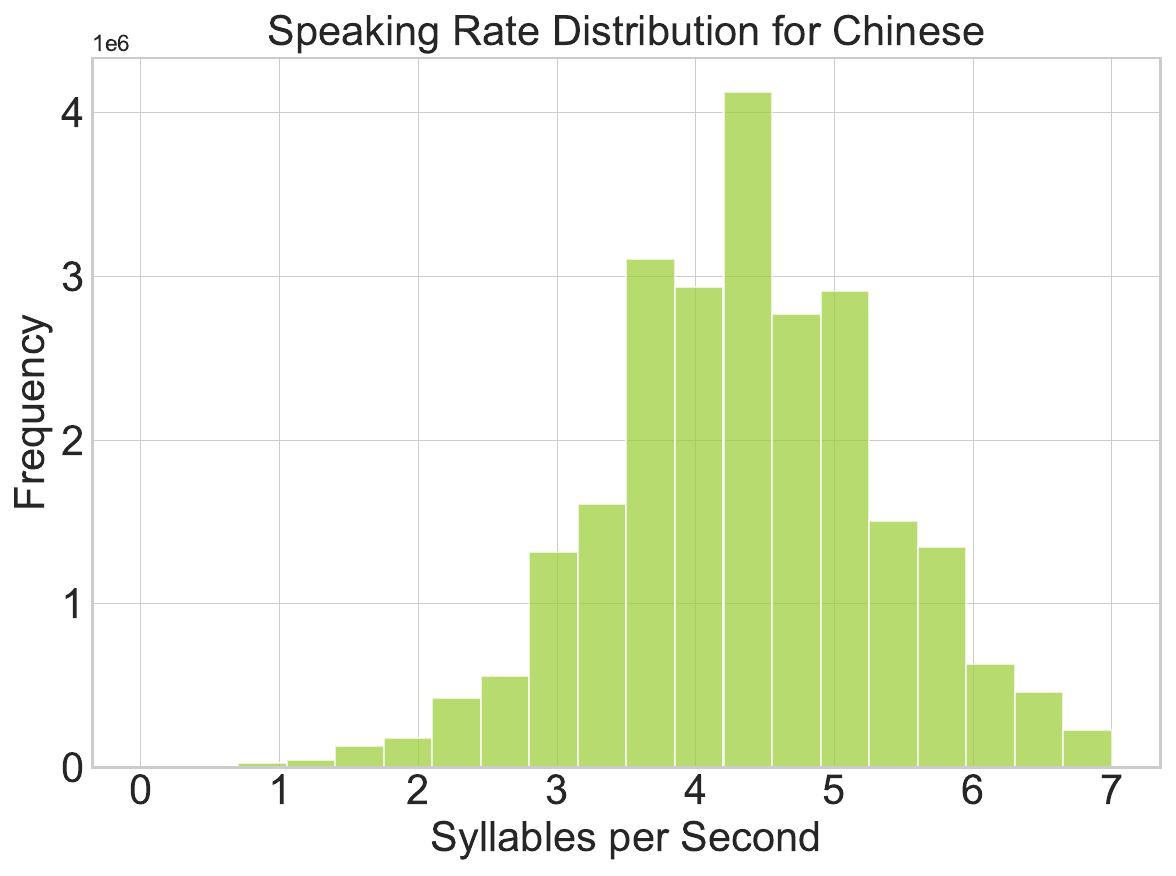}
\end{subfigure}%
\begin{subfigure}{.33\textwidth}
  \centering
  \includegraphics[width=\linewidth]{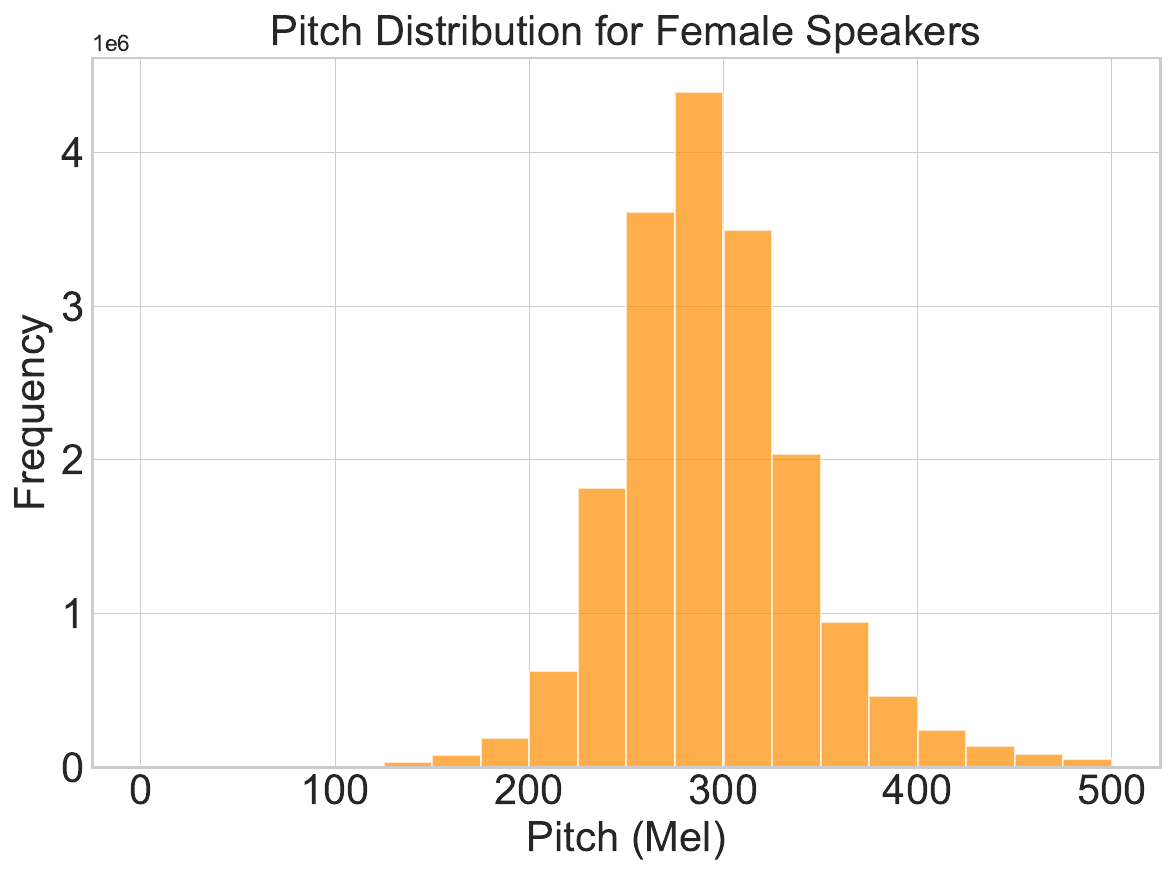}
\end{subfigure}%
\begin{subfigure}{.33\textwidth}
  \centering
  \includegraphics[width=\linewidth]{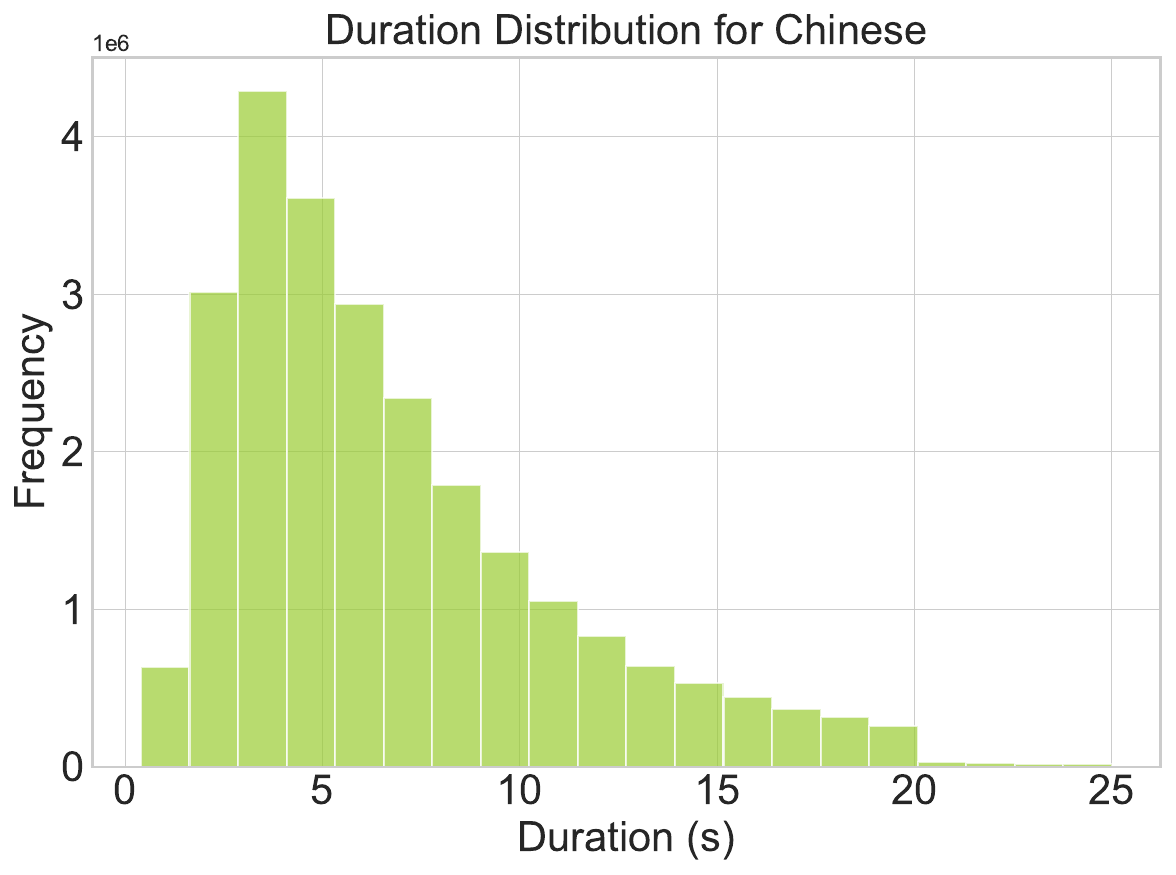}
\end{subfigure}
\caption{Data distribution of VoxBox.}
\label{fig:data_statistic_his}
\end{figure*}

\begin{figure*}[htp]
    \centering
    \includegraphics[width=0.8\linewidth]{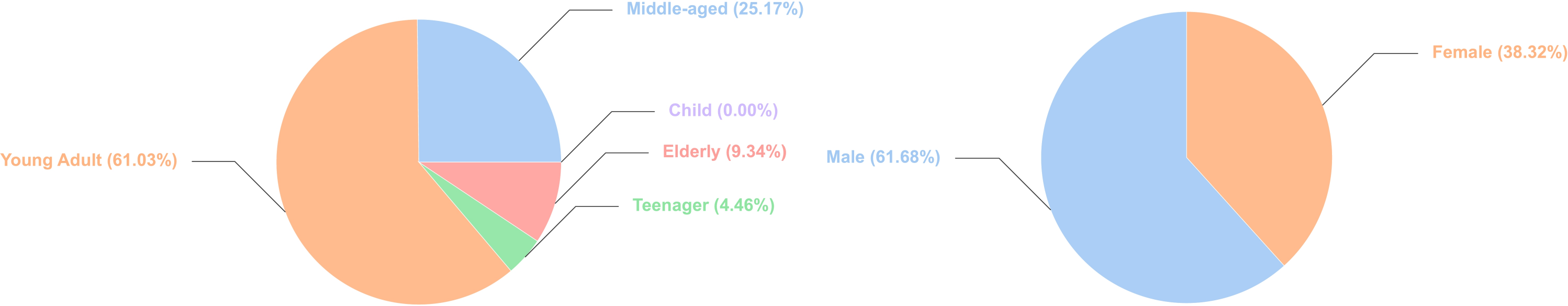}
    \caption{Gender and age distribution of VoxBox.}
    \label{fig:data_statistic_pie}
\end{figure*}

The distributions of speaking rate, duration, and pitch are shown in Fig~\ref{fig:data_statistic_his}, while the distributions of gender and age are presented in Fig~\ref{fig:data_statistic_pie}.

\begin{table*}[tb]
\centering
\caption{VoxBox Statistics}
\setlength{\tabcolsep}{1.mm}
\begin{tabular}{llllll}
\toprule
\multirow{2}{*}{Data} & \multicolumn{1}{c}{\multirow{2}{*}{Language}} & \multicolumn{1}{c}{\multirow{2}{*}{\#Utterance}} & \multicolumn{3}{c}{Duration (h)}         \\ \cline{4-6} 
       & \multicolumn{1}{c}{}           & \multicolumn{1}{c}{}            & \multicolumn{1}{c}{Male} & \multicolumn{1}{c}{Female} & \multicolumn{1}{c}{Total} \\ \midrule
AISHELL-3~\cite{shi2020aishell}             & Chinese         & 88,035           & 16.01     & 69.61       & 85.62      \\
CASIA~\cite{zhang2008design}                & Chinese         & 857              & 0.25      & 0.2         & 0.44       \\
Emilia-CN~\cite{he2024emilia}               & Chinese         & 15,629,241       & 22,017.56 & 12,741.89   & 34,759.45   \\
ESD~\cite{zhou2021seen}                     & Chinese         & 16,101           & 6.69      & 7.68        & 14.37      \\
HQ-Conversations~\cite{zhou2024codec}       & Chinese         & 50,982           & 35.77     & 64.23       & 100        \\
M3ED~\cite{zhao2022m3ed}                    & Chinese         & 253              & 0.04      & 0.06        & 0.1        \\
MAGICDATA~\cite{MAGICDATA2019}              & Chinese         & 609,474          & 360.31    & 393.81      & 754.13     \\
MER2023~\cite{lian2023mer}                  & Chinese         & 1,667            & 0.86      & 1.07        & 1.93       \\
NCSSD-CL-CN~\cite{liu2024generative}        & Chinese         & 98,628           & 53.83     & 59.21       & 113.04     \\
NCSSD-RC-CN~\cite{liu2024generative}        & Chinese         & 21,688           & 7.05      & 22.53       & 29.58      \\
WenetSpeech4TTS~\cite{ma2024wenetspeech4tts}& Chinese         & 8,856,480        & 7,504.19  & 4,264.3     & 11,768.49   \\
Total                                       & Chinese         & 25,373,406       & 30,002.56 & 17,624.59   & 47,627.15   \\ \hline
CREMA-D~\cite{cao2014crema}                 & English         & 809              & 0.3       & 0.27        & 0.57       \\
Dailytalk~\cite{lee2023dailytalk}           & English         & 23,754           & 10.79     & 10.86       & 21.65      \\
Emilia\-EN~\cite{he2024emilia}              & English         & 8,303,103        & 13,724.76 & 6,573.22    & 20,297.98   \\
EMNS~\cite{noriy2023emns}                   & English         & 918              & 0         & 1.49        & 1.49       \\
EmoV-DB~\cite{adigwe2018emotional}          & English         & 3,647            & 2.22      & 2.79        & 5          \\
Expresso~\cite{nguyen2023expresso}          & English         & 11,595           & 5.47      & 5.39        & 10.86      \\
Gigaspeech~\cite{chen2021gigaspeech}        & English         & 6,619,339        & 4,310.19  & 2,885.66    & 7,195.85    \\
Hi-Fi TTS~\cite{bakhturina2021hi}           & English         & 323,911          & 133.31    & 158.38      & 291.68     \\
IEMOCAP~\cite{busso2008iemocap}             & English         & 2,423            & 1.66      & 1.31        & 2.97       \\
JL-Corpus~\cite{james2018open}              & English         & 893              & 0.26      & 0.26        & 0.52       \\
Librispeech~\cite{panayotov2015librispeech} & English         & 230,865          & 393.95    & 367.67      & 761.62     \\
LibriTTS-R~\cite{koizumi2023libritts}       & English         & 363,270          & 277.87    & 283.03      & 560.9      \\
MEAD~\cite{wang2020mead}                    & English         & 3,767            & 2.26      & 2.42        & 4.68       \\
MELD~\cite{poria2018meld}                   & English         & 5,100            & 2.14      & 1.94        & 4.09       \\
MLS-English~\cite{Pratap2020MLSAL}          & English         & 6,319,002        & 14,366.25 & 11,212.92   & 25,579.18   \\
MSP-Podcast~\cite{martinez2020msp}          & English         & 796              & 0.76      & 0.56        & 1.32       \\
NCSSD-CL-EN~\cite{liu2024generative}        & English         & 62,107           & 36.84     & 32.93       & 69.77      \\
NCSSD-RL-EN~\cite{liu2024generative}        & English         & 10,032           & 4.18      & 14.92       & 19.09      \\
RAVDESS~\cite{livingstone2018ryerson}       & English         & 950              & 0.49      & 0.48        & 0.97       \\
SAVEE~\cite{jackson2014surrey}              & English         & 286              & 0.15      & 0.15        & 0.31       \\
TESS~\cite{yu2021vector}                    & English         & 1,956            & 0         & 1.15        & 1.15       \\
VCTK~\cite{yamagishi2019vctk}               & English         & 44,283           & 16.95     & 24.51       & 41.46      \\
Total                                       & English         & 22,332,806       & 33,290.8  & 21,582.31   & 54,873.11   \\ \midrule
\multicolumn{2}{l}{Overall Total}                             & 47,706,212       & 63,293.36 & 39,206.9    & 102,500.26  \\ \bottomrule
\end{tabular}
\label{tab:data}
\end{table*}

\subsection{Source Data}
\label{ap:source_data}

\begin{itemize}
    \item \textbf{AISHELL-3}: A multi-speaker Mandarin speech corpus for TTS.  Source: \url{https://www.openslr.org/93/} 

    \item \textbf{CASIA}: An emotional multi-speaker Mandarin speech corpus containing six emotions for TTS.  Source: \url{https://gitcode.com/open-source-toolkit/bc5e6}

    \item \textbf{CREMA-D}: An emotional multi-speaker multilingual speech corpus containing six emotions and four intensity levels for TTS. Source:\url{https://github.com/CheyneyComputerScience/CREMA-D}

    \item \textbf{Dailytalk}: A multi-speaker English speech corpus with conversational style for TTS. Source:\url{https://github.com/keonlee9420/DailyTalk}

    \item \textbf{Emilia}: A multi-speaker multilingual speech corpus containing six languages for TTS.  Source: \url{https://emilia-dataset.github.io/Emilia-Demo-Page/}

    \item \textbf{EMNS}: An emotional single-speaker English speech corpus for TTS.  Source: \url{https://www.openslr.org/136}

    \item \textbf{EmoV-DB}: An emotional multi-speaker English speech corpus containing four emotions for TTS.  Source: \url{https://mega.nz/folder/KBp32apT#gLIgyWf9iQ-yqnWFUFuUHg/mYwUnI4K}

    \item \textbf{ESD}: An emotional multi-speaker bilingual speech corpus containing five emotions for TTS.  Source: \url{https://hltsingapore.github.io/ESD/}

    \item \textbf{Expresso}: A multi-speaker English speech corpus with reading and improvising conversational style for TTS.  Source: \url{https://speechbot.github.io/expresso/}

    \item \textbf{Gigaspeech}: A multi-speaker English speech corpus with reading  style for TTS.  Source: \url{https://github.com/SpeechColab/GigaSpeech}

    \item \textbf{Hi-Fi TTS}: A multi-speaker English speech corpus with reading  style for TTS.  Source: \url{https://openslr.org/109/}

    \item \textbf{HQ-Conversations}: A mutli-speaker Mandarin speech corpus with conversational style for TTS. Source: \url{https://www.magicdatatech.com/iscslp-2024/}

    \item \textbf{IEMOCAP}: An emotional multi-speaker English speech corpus containing five emotions for TTS.  Source: \url{https://sail.usc.edu/iemocap/iemocap_release.htm}

    \item \textbf{JL-Corpus}: An emotional multi-speaker English speech corpus containing five primary emotions and five secondary emotions for TTS.  Source: \url{https://www.kaggle.com/datasets/tli725/jl-corpus}

    \item \textbf{Librispeech}: A mutli-speaker English speech corpus with reading style for TTS. Source: \url{https://tensorflow.google.cn/datasets/catalog/librispeech}

    \item \textbf{LibriTTS-R}: Sound quality improved version of the LibriTTS~\cite{zen2019libritts} corpus which is a large-scale corpus of English speech for TTS. Source: \url{https://www.openslr.org/141/}

    \item \textbf{M3ED}: An emotional mutli-speaker Mandarin speech corpus containing seven emotions for TTS. Source: \url{https://github.com/aim3-ruc/rucm3ed}

    \item \textbf{MAGICDATA}: A mutli-speaker Mandarin speech corpus with conversational style for TTS. Source: \url{https://openslr.org/68/}

    \item \textbf{MEAD}: An emotional mutli-speaker English speech corpus containing eight emotions and three intensity levels for TTS. Source: \url{https://github.com/uniBruce/Mead}

    \item \textbf{MELD}: An emotional mutli-speaker English speech corpus containing seven emotions for TTS. Source: \url{https://affective-meld.github.io/}

    \item \textbf{MER2023}: An emotional mutli-speaker Mandarin speech corpus containing six emotions for TTS. Source: \url{http://www.merchallenge.cn/datasets}

    \item \textbf{MLS-English}: A  mutli-speaker English speech corpus for TTS. Source: \url{https://www.openslr.org/94/}

    \item \textbf{MSP-Podcast}: An emotional mutli-speaker English speech corpus containing eight emotions for TTS. Source: \url{https://ecs.utdallas.edu/research/researchlabs/msp-lab/MSP-Podcast.html}

    \item \textbf{NCSSD-CL}: A mutli-speaker bilingual speech corpus for TTS. Source: \url{https://github.com/uniBruce/Mead}

    \item \textbf{NCSSD-RL}: A mutli-speaker bilingual speech corpus for TTS. Source: \url{https://github.com/uniBruce/Mead}

    \item \textbf{RAVDESS}: An emotional mutli-speaker English speech corpus containing eight emotions and two intensity levels for TTS. Source: \url{https://www.kaggle.com/datasets/uwrfkaggler/ravdess-emotional-speech-audio}

    \item \textbf{SAVEE}: An emotional mutli-speaker English speech corpus containing seven emotions for TTS. Source: \url{https://www.kaggle.com/datasets/ejlok1/surrey-audiovisual-expressed-emotion-savee}

    \item \textbf{TESS}: An emotional mutli-speaker English speech corpus containing seven emotions for TTS. Source: \url{https://tspace.library.utoronto.ca/handle/1807/24487}

    \item \textbf{VCTK}: A mutli-speaker English speech corpus for TTS. Source: \url{https://datashare.ed.ac.uk/handle/10283/2651}

    \item \textbf{WenetSpeech4TTS}: A large-scale mutli-speaker Mandarin speech corpus for TTS. Source: \url{https://wenetspeech4tts.github.io/wenetspeech4tts/}

\end{itemize}

\section{SparkVox: A Toolkit for Speech Related Tasks}

\begin{figure*}[htp]
    \centering
    \includegraphics[width=\linewidth]{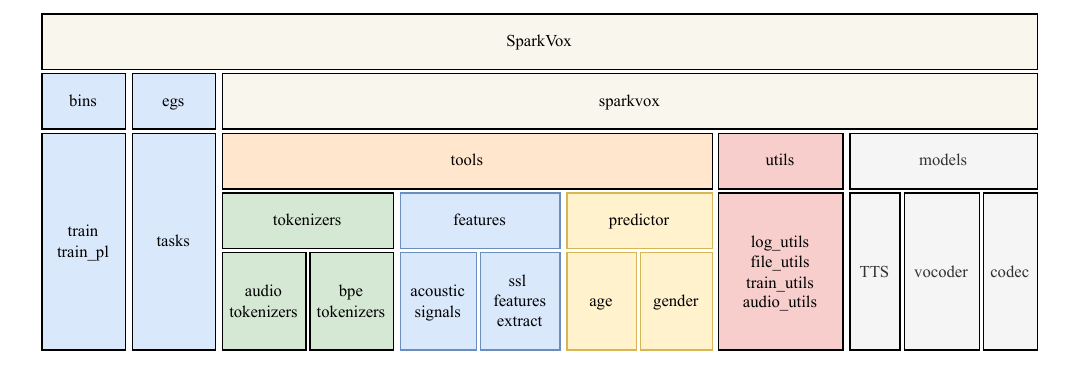}
    \caption{Framework of SparkVox.}
    \label{fig:sparkvox}
\end{figure*}

The training code for Spark-TTS will be integrated into the open-source SparkVox framework. SparkVox is a training framework designed for speech-related tasks, supporting a variety of applications, including: vocoder, codec, TTS, and speech understanding. Additionally, SparkVox provides various file processing tools for both text and speech data, facilitating efficient data handling. Its simplified framework structure is illustrated in Fig.~\ref{fig:sparkvox}.

\end{document}